\documentclass[twocolumn]{aastex63}
\usepackage{amsmath}
\usepackage[caption=false]{subfig}
\usepackage{graphicx}
\usepackage{mathrsfs}

\submitjournal{ApJ}
\shorttitle{Stochastic Photometric Variability in Massive Stars}
\shortauthors{Cantiello et al.}

\def\code#1{\texttt{#1}}

\definecolor{purple}{rgb}{0.5,0,0.5}
\definecolor{darkgreen}{rgb}{0.1,0.6,0.1}
\definecolor{orange}{rgb}{1,0.6,0}

\newcommand{\brvs}{Br\"unt-V\"ais\"al\"a}

\def\hp{\mathrm{H}_{\mathrm{P}}}
\newcommand{\alphamlt}{\alpha_{\rm MLT}}
\newcommand{\nuchar}{\nu_{\rm char}}
\newcommand{\rhoa}{\overline{\rho}_{\rm c}}
\newcommand{\macha}{\overline{\rm M}_{\rm c}}
\newcommand{\vc}{v_{\rm c}}
\newcommand{\vca}{\overline{v}_{\rm c}}
\newcommand{\hpa}{\overline{\mathrm{H}}_{\mathrm P}}
\newcommand{\turnover}{\tau_{\rm c}}
\newcommand{\logl}{\log \mathscr{L}}
\newcommand{\loglsun}{\log \mathscr{L}/\mathscr{L}_\odot}
\newcommand{\logteff}{\log {\rm T}_{\rm eff}}
\newcommand{\kms}{\rm km s^{-1}}
\newcommand{\muhz}{\mu\textrm{Hz}}
\newcommand{\nucharobs}{\nu_{\rm char}}
\newcommand{\alphazero}{\alpha_0}
\newcommand{\mso}{{\rm M}_\odot}

 \def\Rtop{\mathrm{R}_{\mathrm{c}}}

\begin{document}
\title{On the Origin of Stochastic, Low-Frequency Photometric Variability in Massive Stars}

\author[0000-0002-8171-8596]{Matteo Cantiello}
\affiliation{Center for Computational Astrophysics, Flatiron Institute, New York, NY 10010, USA}
\affiliation{Department of Astrophysical Sciences, Princeton University, Princeton, NJ 08544, USA}

\author[0000-0002-7635-9728]{Daniel Lecoanet}
\affiliation{Department of Engineering Sciences \& Applied Mathematics, Northwestern University, Evanston, IL 60201, USA}
\affiliation{CIERA, Northwestern University, Evanston, IL 60201, USA}
\affiliation{Department of Astrophysical Sciences, Princeton University, Princeton, NJ 08544, USA}

\author[0000-0001-5048-9973]{Adam S. Jermyn}
\affiliation{Center for Computational Astrophysics, Flatiron Institute, New York, NY 10010, USA}

\author[0000-0001-9324-3359]{Luca Grassitelli}
\affiliation{Argelander-Institut f\"ur Astronomie der Universit\"at Bonn, Auf dem H\"ugel 71, D-53121 Bonn, Germany}

\correspondingauthor{Matteo Cantiello}
\email{mcantiello@flatironinstitute.org}

\begin{abstract}
High-precision photometric observations have revealed ubiquitous stochastic low-frequency photometric variability in early type stars.
It has been suggested that this variability arises due to either subsurface convection or internal gravity waves launched by the convective core.
Here we show that relevant properties of convection in subsurface convective layers correlate very well with the timescale and amplitude of stochastic low-frequency photometric variability, as well as with the amplitude of macroturbulence.
We suggest that low-frequency, stochastic photometric variability and surface turbulence in massive stars are caused by the the presence of subsurface convection.
We show that an explanation for the observed surface photometric variability and macroturbulence relying on convective core driven internal gravity waves encounters a number of difficulties and seems unlikely to be able to explain the observed trends.

\end{abstract}


\section{Introduction}\label{sec:introduction}
Massive stars largely drive the dynamical and chemical evolution of gas in galaxies \citep[e.g.][]{Hopkins:2014}. They accomplish this via their stellar winds, eruptions, and explosive deaths, ultimately producing neutron stars and black holes \citep{Langer:2012}.
These compact remnants can merge and generate the gravitational waves observed by LIGO/Virgo \citep{2016PhRvL.116f1102A}. The evolutionary trajectory starting with a massive star burning hydrogen in its core and ending with a compact remnant is understood only qualitatively. We still do not know how to map initial properties of the star, like mass, rotation rate, and metallicity, to e.g. the final mass and spin of the compact remnant it leaves behind. The picture is further complicated by the fact that the majority of massive stars are found in multiple systems \citep{Sana:2012}, with a large fraction expected to interact with their companions \citep{deMink:2014}. 

The path towards a detailed understanding of massive stars begins with a quantitative study of their internal structure on the main sequence. In recent years asteroseismology has opened a new window on these challenging astrophysical environments, with high precision photometry from space delivering many new exciting results \citep[e.g., MOST, CoRoT, BRITE, Kepler/K2 and TESS, see ][]{BowmanReview:2020}.
The latest discovery is the detection of a new ubiquitous phenomenon in massive stars: stochastic low-frequency photometric variability \citep[SLF variability;][]{Blomme:2011,2019NatAs...3..760B,Bowman:2019,Pedersen:2019,Bowman:2020,Rauw:2021}. 
This joins a number of other surface and wind phenomena that are routinely observed in early-type stars, including surface velocity fluctuations  \citep[Macroturbulence;][]{Simon-Diaz:2014}, line profile variability \citep{Fullerton:1996}, and discrete absorption components in UV spectra \citep{Howarth:1989,Cranmer:1996,Fullerton:1997,Kaper:1997}. Surface magnetism and bright spots are harder to observe but could still be common in these stars \citep[e.g.][]{Ramiaramanantsoa:2014}.

The origin of this SLF variability is currently debated.
It could be caused by sub-surface convection zones \citep{Cantiello:2009,Blomme:2011,Lecoanet:2020} or by internal gravity waves (IGWs) launched by turbulent core convection \citep{Edelmann:2019,Ratnasingam:2020}\footnote{Classical heat-mechanism pulsations could be responsible for
spectroscopic and photometric variability in specific parts of the HRD, but
they can hardly justify the apparent ubiquity of macroturbulence and SLF in
massive stars \citep{Godart:2017,Simon-Diaz:2017}.}. Instabilities in the stellar wind could also play a role \citep{Krticka:2021}.
Regardless of its origin, this photometric signal likely carries important information about stellar structure, complementing asteroseismic studies that use well-identified oscillation modes \citep[e.g.][]{Aerts:2019,Aerts_araa:2019,BowmanReview:2020}. 

Recently the use of high resolution ground-based spectroscopy for the targets observed by K2 and TESS \citep{Bowman:2020} has allowed precise determination of stellar parameters, including spectroscopic mass, luminosity, and macroturbulence \citep{Burssens:2020}.
The latter is particularly important if, as seems likely, the mechanism exciting surface turbulent velocities is the same as that which produces the observed SLF variability \citep{Grassitelli:2016,Bowman:2020}.

Here we combine spectroscopic and photometric data to compare the observed properties of the stochastic photometric variability and macroturbulence with predictions from  non-rotating 1D stellar models. 
We make simple predictions for the amplitude and frequency of the variability that subsurface convection induces at the stellar surface and examine how these vary with stellar temperature and luminosity.
We find that the predicted trends coming from a suburface convection zone driven by the iron opacity peak (FeCZ) at $\approx150$kK match the observations well. 

We next show that one way to differentiate between the two proposed mechanisms is by examining macroturbulence and SLF in massive stars with surface magnetic fields, since magnetic effects have a larger impact on core IGWs than on the FeCZ.
Macroturbulence is observed in stars with fairly strong magnetic fields, sufficient to suppress IGWs from the core, favoring a model based on subsurface convection.
Furthermore, the only stars with no observed macroturbulence are ones where the magnetic field is strong enough to shut off the FeCZ, and so far as we know all stars with such strong magnetic fields lack macroturbulence, consistent with a subsurface origin of surface perturbations~\citep{Jermyn:2020}.

Based on these considerations we suggest that subsurface convection represents a possible unifying mechanism causing SLF variability, surface turbulence, and magnetic spots in massive stars.

\begin{figure}[ht!]
\begin{center}
\includegraphics[width=1.0\columnwidth]{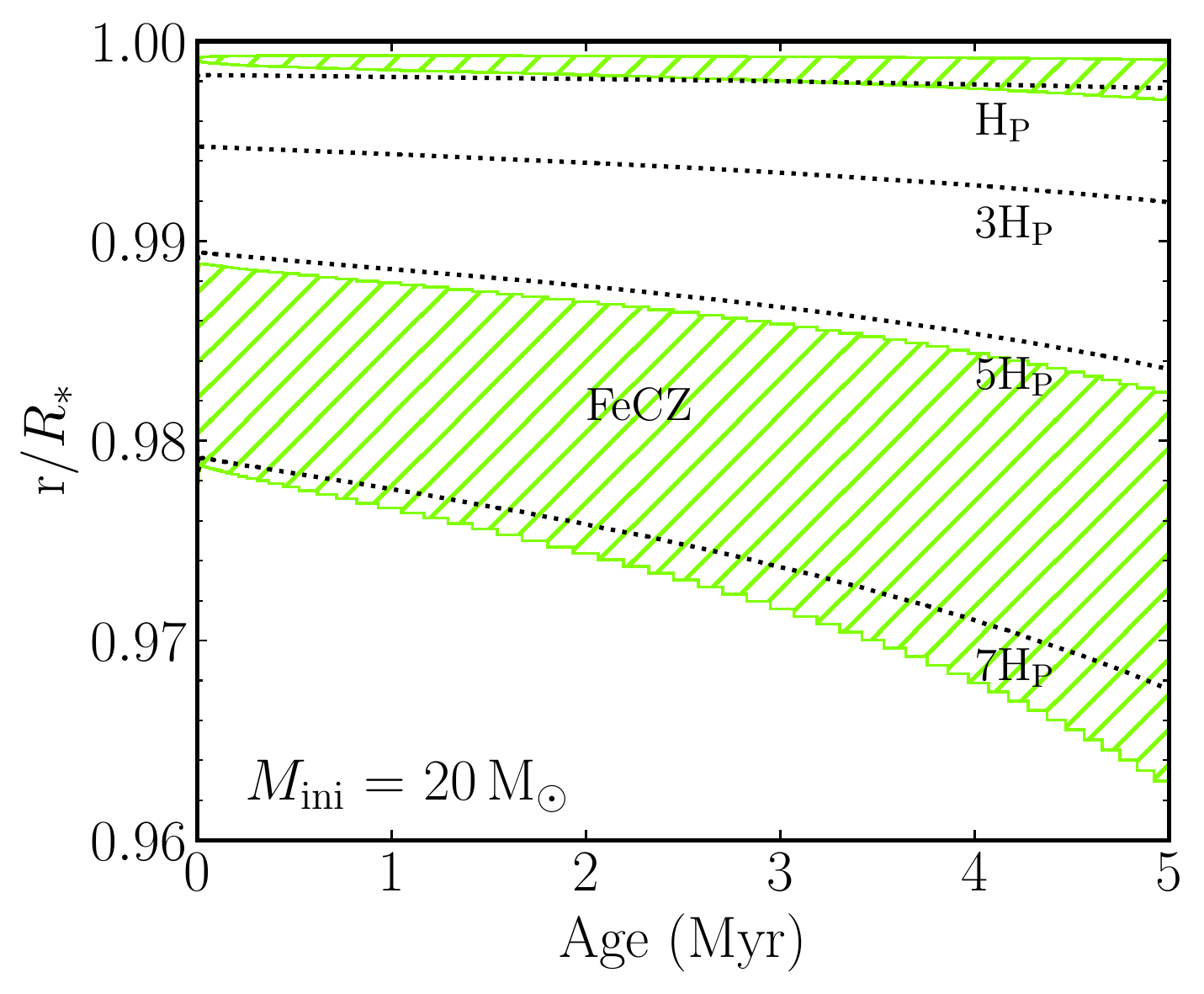}
\caption{\label{fig:kipp20} Evolution of the normalized radial location of the FeCZ in a 20$\mso$ model during the main sequence (first 5 Myr of evolution). The HeCZ is also visible very close to the surface. Locations of P(r)/P($R_*$) = $e^n$  for $n=[1,3,5,7]$ are also shown and labelled as $n {\rm H}_{\rm P}$.}
\end{center}
\end{figure}

\section{Methods}
We calculated stellar evolution models using the Modules for Experiments in Stellar Astrophysics
\citep[MESA][]{Paxton2011, Paxton2013, Paxton2015, Paxton2018, Paxton2019} software instrument.
Details on the microphysics inputs to this software instrument are given in Appendix~\ref{sec:software}. 
 Our models have initial mass ranging from 5 to 120$\mso$ and are non-rotating. Since OB stars are known to be rapidly rotating \citep[e.g.][]{Maeder:2000,Dufton:2013,Ramirez:2013}, we discuss the potential impact of rotation on our results in Section~\ref{sec:rotation}. We also neglect the effect of wind mass-loss. While this process is important for the evolution of massive stars \citep{Smith:2014}, the conditions in  subsurface convective zones and the properties of convection depend almost exclusively from the location in the Hertzsprung-Russell diagram.  This is true as long as the outer layers composition is not substantially altered.
We focused on the structure and convective properties of our models in order to calculate typical frequencies and convective fluxes. 
Convection is calculated in the framework of the Mixing Length Theory \citep[][MLT hereafter]{1958ZA.....46..108B}, and we adopted $\alphamlt = 1.6$. While the properties of efficient convection zones (e.g. stellar cores) are insensitive to the choice of this parameter, those of inefficient convective regions close to the stellar surface \emph{do} depend on $\alphamlt$ \citep{2015ApJ...813...74J,Cantiello:2019}. In Section~\ref{sec:discussion} we discuss how uncertainties in the choice of the $\alphamlt$  parameter affect our results. 

Since we are interested in the excitation of surface phenomena we focus on the upper part of convective zones.
Following \citet{Cantiello:2009}  we define the average of a generic quantity $q$ as
\begin{equation}
\overline{q} \equiv \frac{1}{\alphamlt \hp} \int_{\Rtop-\alphamlt\hp}^{\Rtop} q(r)\; dr,
\label{aver}
\end{equation}
where  $\hp$ is the pressure scale height calculated at the upper boundary ($\Rtop$) of the 
convective zone of interest.  We tested a variety of different average prescriptions and found that our results do not depend much on the specific choice of prescription.  

Using equation~\eqref{aver} we extracted the average Mach number $\macha$, convective velocity $\vca$ and density $\rhoa$ in the convective core and in the subsurface convection zones of our models.
The most important subsurface convection zone for the massive stars we focus on is the FeCZ (see Fig.~\ref{fig:kipp20}), although in the low luminosity regime He-driven convection zones could play a role as well \citep{Cantiello:2009,Cantiello:2019}.

We compare relevant properties of our theoretical models  with the observed characteristic frequency $\nucharobs$ and amplitude $\alphazero$ of SLF variability. These quantities are derived by fitting the stochastically-excited, low-frequency power excess in a power
density spectrum using a Lorentzian function \citep[e.g.][]{Bowman:2019}:

\begin{equation}\label{eq:rednoise}
\alpha(\nu) = \frac{\alphazero}{1+\big(\frac{\nu}{\nucharobs}\big)^{\gamma}} \, + P_{\rm W}.
\end{equation}
This shows that $\alphazero$ represents the amplitude at zero
frequency. $\nucharobs$ is defined as in eqn.~\ref{eq:nuchar} with $\tau$ the typical timescale of the SLF variability.  
Finally, $\gamma$ is the gradient of the linear part of the profile and $P_{\rm W}$ is a white noise term.
Due to its stochastic, low-frequency manifestation in the power density spectrum, in the literature the SLF variability is also referred to as ``red noise'' \citep[e.g.][]{Blomme:2011}.

\section{Results}
We want to test a possible correlation between the properties of the FeCZ and observed photospheric phenomena, in particular SLF variability and turbulent velocities at the stellar surface (macroturbulence).
We proceed by calculating quantities that measure the amplitude of perturbations in the FeCZ. We then check if some of these properties correlate with the amplitude of observed photospheric phenomena, including turbulent velocity fluctuations and SLF variability.  

We define the characteristic frequency as 
\begin{equation}\label{eq:nuchar}
\nuchar \equiv \frac{1}{2\pi\tau},
\end{equation}
where $\tau$ is a characteristic timescale.
For  comparing with observations we set $\tau$ to be the average convective turnover time 
\begin{equation}\label{eq:turnover}
\turnover = \alphamlt \hpa / \vca,
\end{equation}
calculated either in the FeCZ or in the convective core.  

We calculated the convective flux $F_{\rm c} = \rhoa \vca^3$, where $\rhoa$ and $\vca$ are the average density and convective velocity calculated according to eqn.~\ref{aver}. 
We did this as a function of both mass and evolutionary history for stars with initial mass ranging from 5 to 120$~M_\odot$.
Here we present results for an initial metallicity of Z=0.02, but in Appendix~\ref{appen:grids} we report results for model grids with Z=0.006 and 0.002 as well.

\subsection{FeCZ and Macroturbulence}

\begin{figure*}
  \centering
    \subfloat{\includegraphics[width=1\columnwidth]{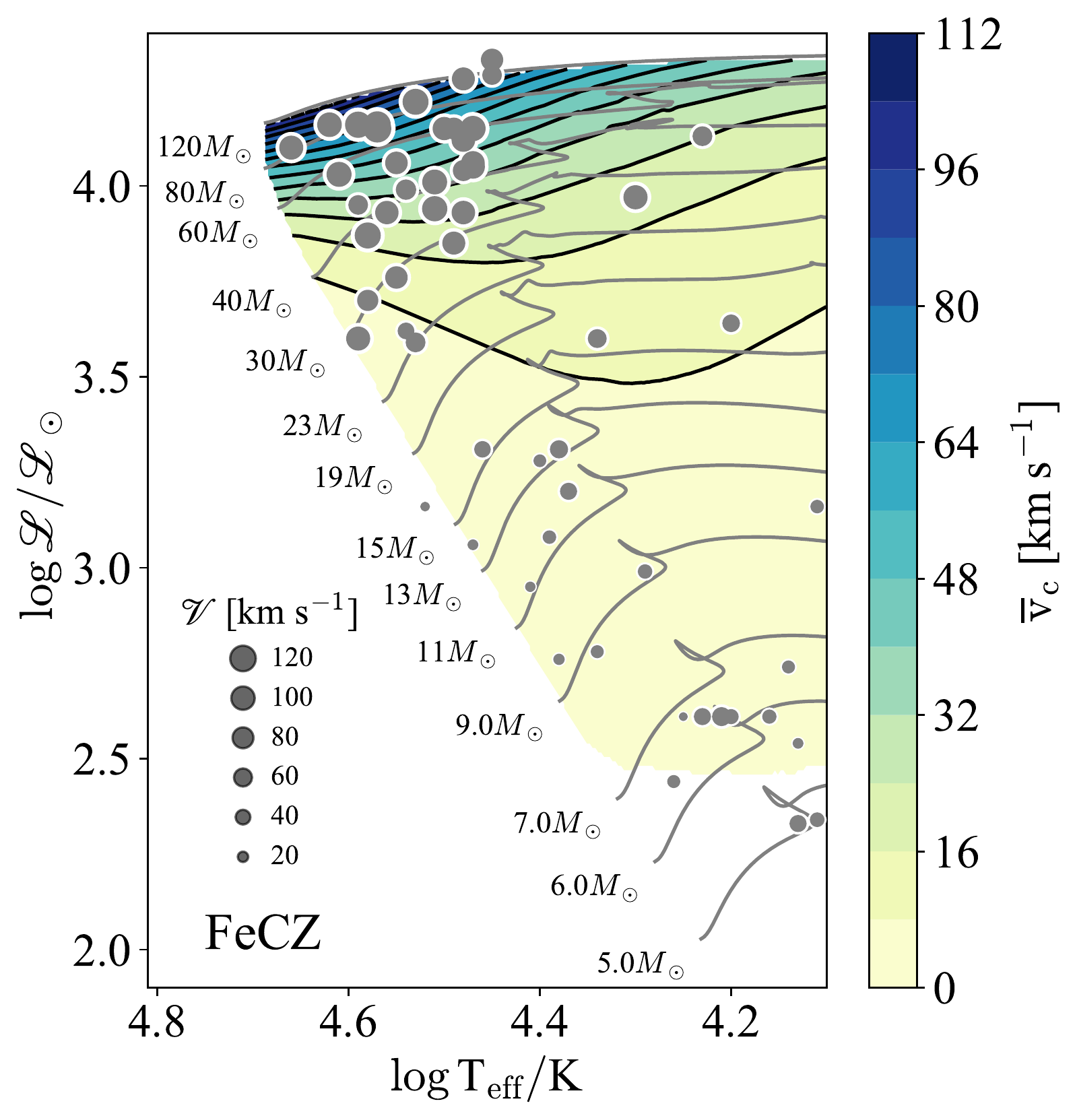}}\hfill
  \subfloat{\includegraphics[width=1.005\columnwidth]{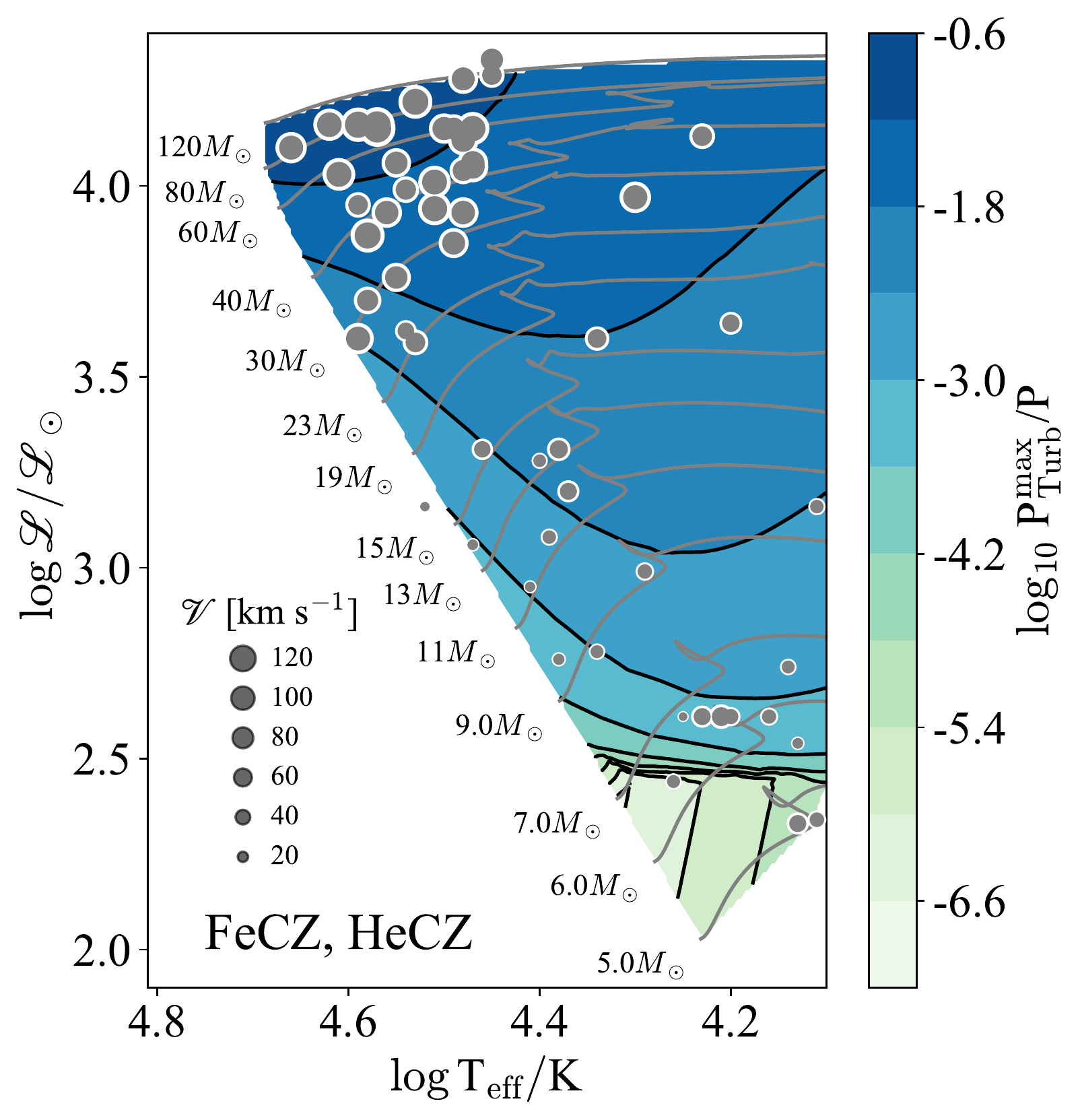}}
 \caption{\label{fig:vconv} Left: Average convective velocities in the FeCZ as a function of the location of stellar models in the spectroscopic H-R Diagram ($\mathscr{L} \equiv {\rm T}^4_{\rm eff}/g$). We also show observed stars with detected macroturbulent velocity as grey circles. The area of the circle is proportional to the observed macroturbulence \citep[data from][]{Burssens:2020,Bowman:2020}.  The FeCZ is absent in models with $\loglsun \lesssim 2.5$. Right: Same as left, but showing predictions for the maximum of the ratio between turbulent pressure and total pressure in any subsurface convection zone. The FeCZ largely dominates, except for stars at low luminosities where turbulent pressure is provided by a helium convection zone (HeCZ).}\label{fig:macroturbulence}
 \end{figure*}

Macroturbulence is a spectroscopic measure of  velocity fields with a scale larger than the photons mean free path in the stellar atmosphere. The shape of  spectral lines can be used not only to measure the amplitude of the velocity field, but also to infer some of its directionality.  
Note that \citet{Simon-Diaz:2014} showed that the line profiles are fitted better by a radial-tangential velocity function  than a Gaussian one, but the observations do not tell if the dominant velocity component is radial or horizontal\footnote{The claim is that the line profiles are fitted better by the radial-tangential velocity function, compared to an isotropic one, with either $v_{\rm r} \gg v_{\rm t}$ or $v_{\rm r} \ll v_{\rm t}$, where $v_{\rm r}$ and $v_{\rm t}$ are the radial and  tangential  components of the velocity, respectively (Sim\'on-D\'iaz, private comm.). 
Below we compare the observed amplitude of macroturbulent velocities with predicted properties of the FeCZ.}

We show the average convective velocity in the FeCZ in Fig.~\ref{fig:vconv}, left panel. Velocities of the order $10 \dots 100\ \kms$ are found across the upper spectroscopic H-R diagram \citep[$\mathscr{L} \equiv {\rm T}^4_{\rm eff}/g$;][]{2014A&A...564A..52L}, with a trend of increasing $\vc$ for higher luminosities. Note that, contrary to core convection, the FeCZ is just mildly subsonic, with Mach numbers ranging from 0.01 to 0.3, though these are uncertain by a factor of $\approx 8$ due to a dependence on the uncertain $\alpha_{\rm MLT}$ \citep{Cantiello:2019}. 

In the right panel of the same figure we also show the maximum of the ratio of turbulent pressure to total pressure (P$_{\rm turb} \propto \vc^2$),  which, in agreement with the results of \citet{Grassitelli:2015}, shows a strong correlation with the spectroscopically-derived macroturbulent velocities in massive stars. 

The two quantities in Fig.~\ref{fig:vconv} measure the strength of the inefficient convection. 
We do not yet know the exact mechanism connecting the FeCZ with the surface velocity perturbations.
If one assumes convective elements conserve their inertia as they reach layers stable against convection \citep[e.g., because they are thermally diffusive, see][]{2015ApJ...813...74J}, then the surface velocities should be proportional to the convective velocity; see the left panel of Fig.~\ref{fig:vconv}.
Alternatively, one can assume IGWs are excited with pressure perturbations $\delta p \sim P_{\rm turb}\propto \vc^2$, the turbulent pressure of the convection \citep{Press:1981}.
This stochastic excitation can lead to both running waves and standing modes. Using the polarization relations of adiabatic IGWs \citep[e.g.,][]{sutherland_2010}, a wave with pressure perturbation $\delta p$ has an associated horizontal velocity $u_h \sim (\delta p/\rho_0) k_h/\omega$, where $k_h$ is the horizontal wavenumber of the wave and $\omega$ is its frequency.
The dominant waves will have $k_h\sim1/{\mathrm{H}}_{\mathrm P}$ and $\omega\sim1/\tau_c$ \citep{Cantiello:2009}.
So running IGWs would have surface velocities which also scale like $u_h \sim \vc$ at the surface.
Concerning mode excitation, \citet{Grassitelli:2015} argue that the ratio of turbulent pressure to total pressure in the FeCZ traces the stochastic Lagrangian pressure perturbation associated with the convective motions.  This is responsible for local deviation from hydrostatic equilibrium and the excitation of high-order pulsations with frequencies close to the spectrum of the fluctuations. The ratio of the turbulent to total pressure in the FeCZ is reported in the right panel of Fig.~\ref{fig:vconv}.  

We confirm that convective velocities and the ratio of turbulent pressure to total pressure in the FeCZ correlate very well with the amplitude of macroturbulence.

\subsection{FeCZ and Stochastic, Low-Frequency Variability}

\begin{figure*}[htp]
  \centering
  \subfloat{\includegraphics[width=1.0\columnwidth]{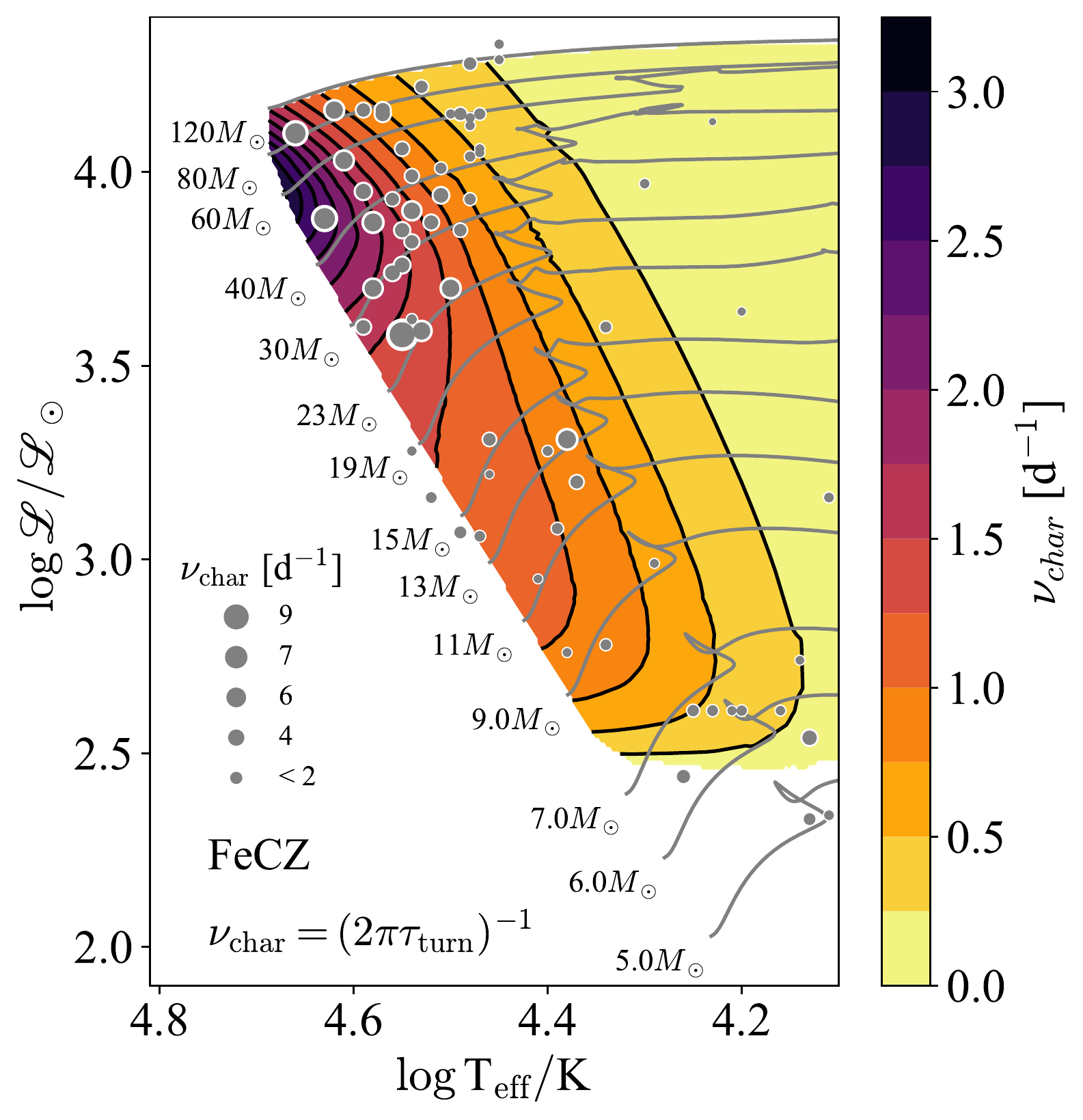}}\hfill
    \subfloat{\includegraphics[width=1\columnwidth]{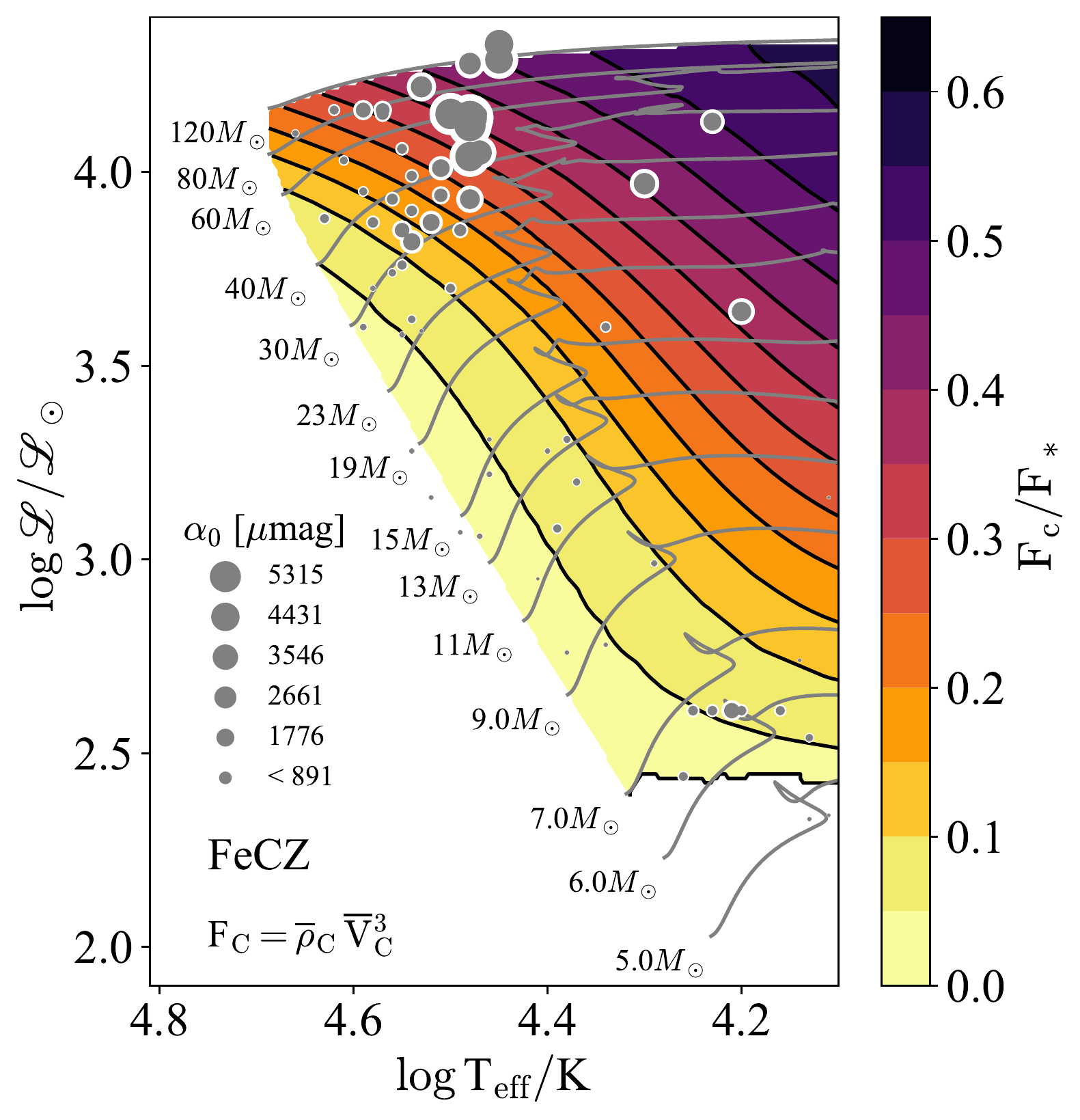}}
  
   \caption{  Left panel:   Characteristic frequency $\nu_{\rm char}$ in the FeCZ as a function of the location of stellar models in the spectroscopic H-R Diagram (black contour lines). Evolutionary tracks are shown as grey solid lines. We also show the observed stars with detected stochastic photometric variability as grey circles. The area of the circle is proportional to the observed $\nu_{\rm char}$, derived from eqn.~\eqref{eq:rednoise} fitting the data in the range $0.1 \le
\nu \le 360\,$d$^{-1}$ \citep{Burssens:2020,Bowman:2020}. Right panel: Ratio of FeCZ convective flux to the total stellar flux in the spectroscopic H-R Diagram ($\mathscr{L} \equiv {\rm T}^4_{\rm eff}/g$). We also show the observed stars with detected stochastic, low-frequency photometric variability as grey circles. The area of the circle is proportional to $\alpha_0$.}
    \label{fig:nuchar_fecz}
\end{figure*}

If the SLF variability is caused by the FeCZ, a natural choice of proxy for the typical timescale of this variability is the convective turnover timescale.  We find typical values of the convective turnover timescale to be about $\sim 0.1\dots2$ d in the FeCZ, with a tendency for shorter values in models with high effective temperature and surface luminosity \citep{Cantiello:2009}.
We computed $\nuchar$ using eqn.~\ref{eq:nuchar} and plot this alongside observed characteristic frequencies\footnote{These were inferred via eqn.~\eqref{eq:rednoise}.} of SLF variability on the spectroscopic H-R diagram, see left panel in Fig.~\ref{fig:nuchar_fecz}.

We see good agreement, with our models reproducing both the typical values of the observed $\nuchar$ and the trend with  $\logl$ and $\logteff$. 
While the characteristic frequencies found by \citet{Bowman:2020} seem to be larger by a factor of $\approx$ 3, our predictions for the turnover timescale are affected by uncertainty in the convective velocities arising from the MLT treatment ($\vc \propto \alphamlt^{3}$, so $\nuchar \propto \alphamlt^{2}$),  as well as our limited knowledge of the frequency spectrum generated by turbulent convection. Since $\alphamlt$ is uncertain by a factor of 2 or so, our estimates of the characteristic frequency in the FeCZ are uncertain by a factor of  $\approx$ 4 and so are consistent with the observations.

We also compare the amplitude of SLF variability with the relative convective flux in the FeCZ (right panel in Fig.~\ref{fig:nuchar_fecz}). Values of F$_{\rm c}$/F$_{*}$ tend to increase with increasing $\logl$ and decreasing $\logteff$, a trend that is also found for the amplitude of observed SLF variability. 

Overall the turnover timescale and relative flux in the FeCZ correlate very well with the observed timescale and amplitude of SLF variability. 
 
\subsection{Core Convection and Stochastic, Low-Frequency Variability}

\begin{figure*}[htp]
  \centering
  
      \subfloat{\includegraphics[width=1\columnwidth]{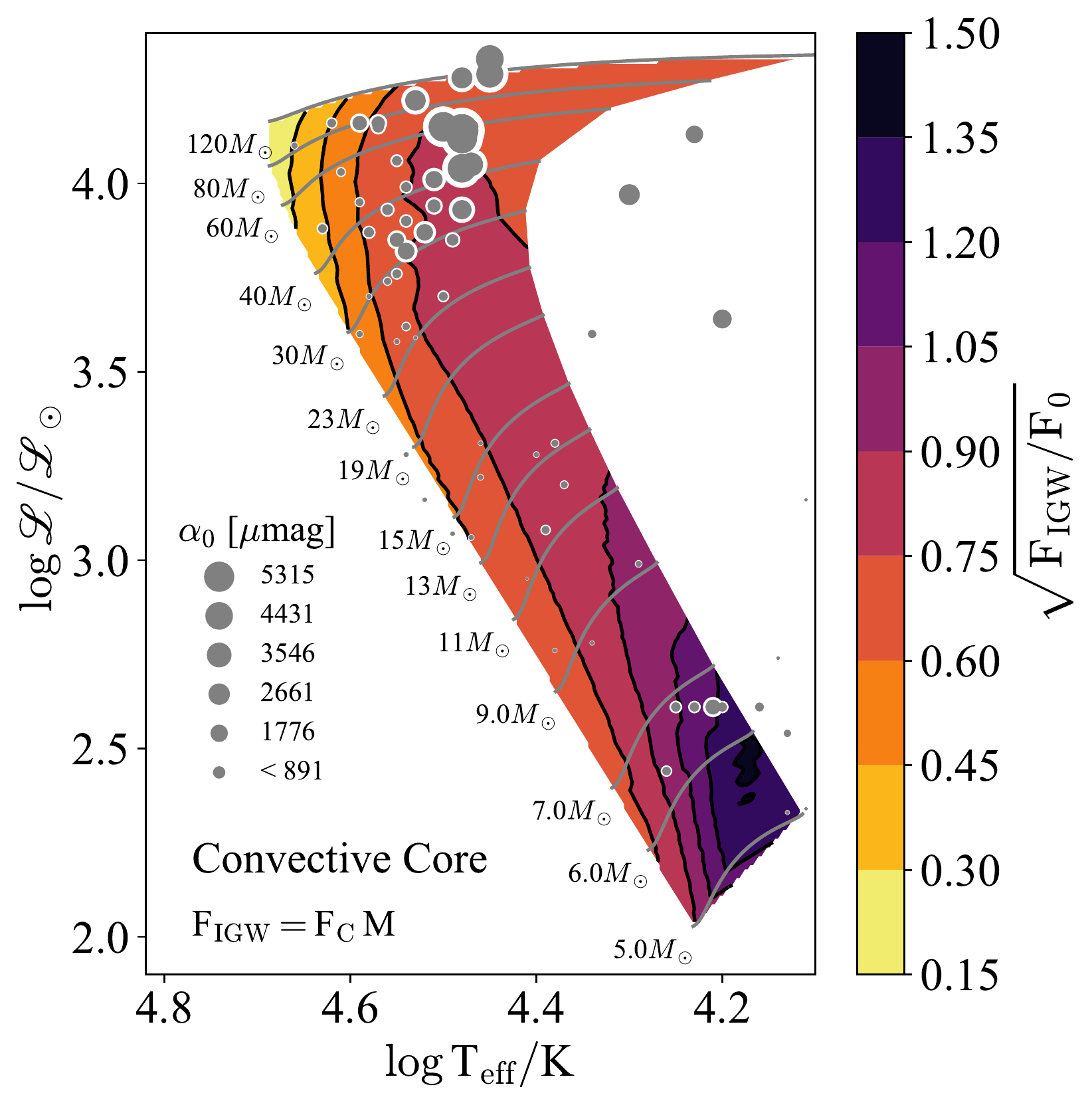}}\hfill
      \subfloat{\includegraphics[width=1.03\columnwidth]{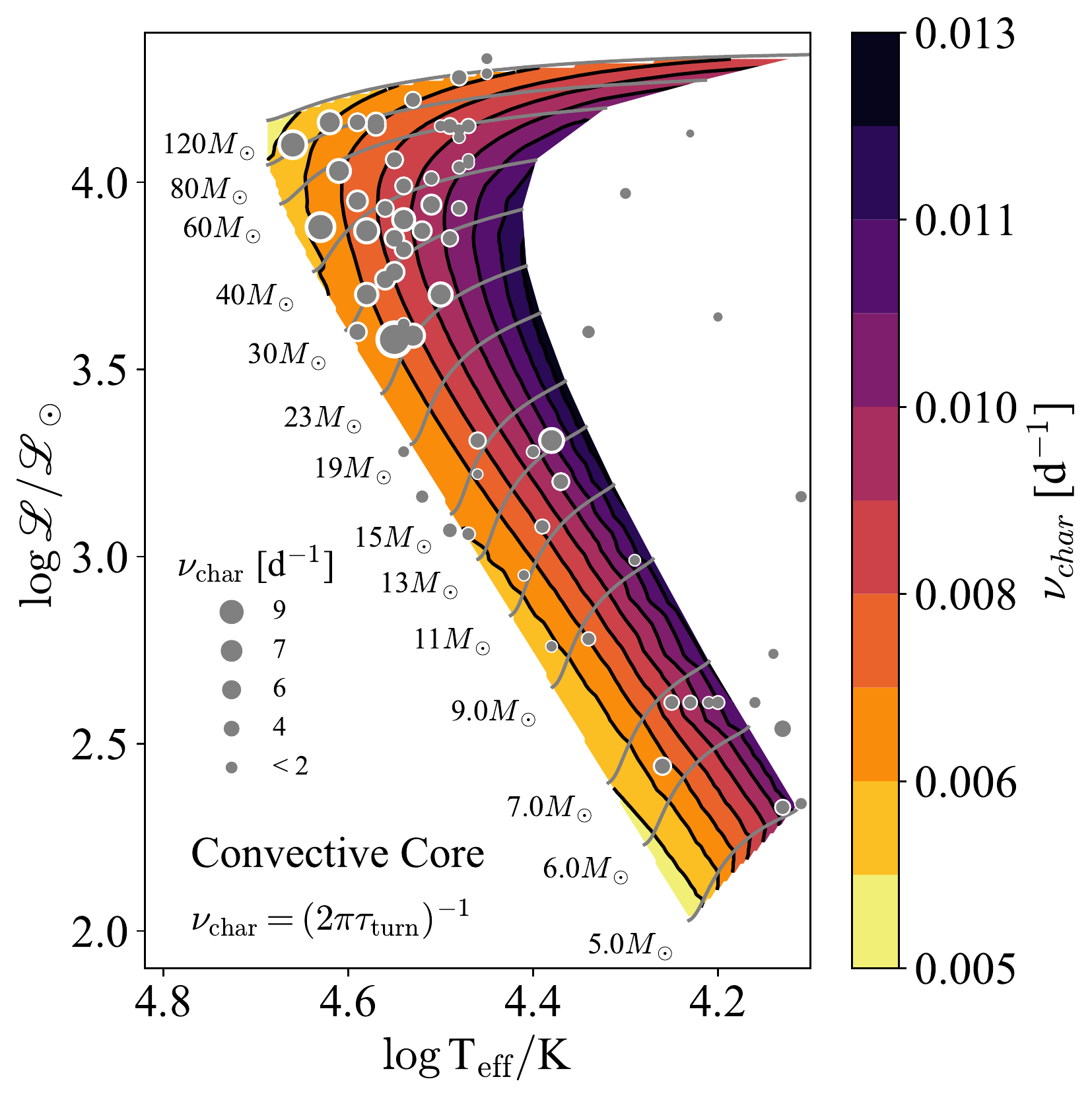}}

\caption{ Left panel:  Square root of the ratio between the flux of gravity waves launched by the convective core and $F_0 = 1/2 \rho(r) \,r^3 \omega^2 \sqrt{N(r)^2-\omega^2}$, evaluated at the stellar surface. In the absence of damping, this quantity is expected to be proportional to the relative radial displacement (Appendix~\ref{appen:xir}) and hence to the relative luminosity fluctuations at the surface.  The IGW flux is calculated multiplying the core convective flux by the average convective Mach number in the top pressure scale height of the convective region. We also show the observed stars with detected stochastic, low-frequency photometric variability as grey circles. The area of the circle is proportional to $\alpha_0$, derived from eqn.~\eqref{eq:rednoise} fitting the data in the range $0.1 \le
\nu \le 360\,$d$^{-1}$  \citep{Burssens:2020,Bowman:2020}.
 Right panel: Characteristic frequency $\nu_{\rm char}$ in the convective core as function of the location of stellar models in the spectroscopic H-R diagram. We also show the observed stars with detected photometric variability as grey circles. The area of the circle is proportional to $\nu_{\rm char}$.  }
    \label{fig:nuchar_core}
\end{figure*}

Early-type stars posses convective cores during their main sequence. Turbulent convection in these convective cores excites internal gravity waves \citep{1990ApJ...363..694G,2013ApJ...772...21R,2013MNRAS.430.2363L,Shiode:2013,Edelmann:2019,Horst:2020}, that can propagate through the stellar envelope and reach the stellar surface \citep[e.g.][]{Ratnasingam:2019,Lecoanet:2020,Ratnasingam:2020}.
Some groups have argued that such waves could be responsible for both the observed macroturbulence and SLF variability in early-type stars \citep{2009A&A...508..409A,2017A&A...597A..22S,Bowman:2019,2019NatAs...3..760B}.
 
There is substantial uncertainty in the surface brightness fluctuations from IGWs generated by core convection. \citet{Shiode:2013} predicted the typical amplitude of g-modes excited by convection to be $\approx 10^{-2}-10^2 \, \mu{\rm mag}$, seemingly at odds with the relatively large amplitudes $\approx 10-10^4 \mu{\rm mag}$ observed by e.g.~\citet{Bowman:2020}. However, \citet{Lecoanet:2021} recently found an error in the \citet{Shiode:2013} prediction. The g-mode amplitude should be larger by a factor of $\approx \sqrt{\nu/\gamma}$, where $\gamma$ is the mode's damping rate. This missing factor could increase the predicted g-mode amplitude by a factor of $10^4$ or larger for high-frequency modes near the Brunt-V\"ais\"al\"a frequency (e.g., $\sim 10\, d^{-1}$ for a $10 M_\odot$ ZAMS star), but does not change the predicted mode amplitudes for lower frequency waves (e.g., $\sim 0.3\, d^{-1}$ for a $10 M_\odot$ ZAMS star). \citet{Lecoanet:2020} argued that there should be very low wave power at low frequencies due to radiative damping, while the wave signal at frequencies above $0.5\, d^{-1}$ should be dominated by g-modes, as predicted by \citet{Shiode:2013}. These features do not seem to be present in the observed spectra.

 Recent numerical simulations of wave generation by convection in a $3M_\odot$ star \citep{Edelmann:2019,Horst:2020} produce wave fluctuation spectra which are qualitatively similar to those observed. However, those simulations artificially boost the stellar luminosity by factors ranging from $10^3$ to $10^7$.  Boosting the luminosity should both enhance the wave amplitude and increase the typical frequency of excited waves, making it difficult to quantitatively compare to observations. 

Although the detailed physics of wave generation by convection is uncertain, we can still analyze the properties of core convective of our models.
If the surface variability is due to core convection, one would expect the characteristic frequency and amplitude of the SLF variability to correlate with the properties of the core convection.
The flux of IGWs excited by turbulent convection at the core boundary is of order F$_{\rm IGW} = {\rm F}_* \macha$ \citep{1990ApJ...363..694G},  where we evaluate the average Mach number $\macha$ using eqn.~\ref{aver} and the local adiabatic sound speed.  We expect the luminosity fluctuations to be proportional to the relative surface radial displacement $\xi_r/R$ produced by these waves at the stellar surface \citep[e.g.][]{Dziembowski:1977,Aerts:2010}.  It can be shown that  $\xi_r/R \propto \sqrt{{\rm F}_{\rm IGW}/ {\rm F}_0} $ (see Appendix~\ref{appen:xir}).  This quantity is presented in Fig.~\ref{fig:nuchar_core} along with the observed amplitudes of SFL variability. 

One important caveat is that this estimate neglects the important role of radiative damping, which is essential in shaping both the amplitude and the shape of the spectrum of waves at the surface \citep[e.g.][]{Lecoanet:2020}. Rotation might also be key in setting the amplitude and shape of the surface fluctuations spectrum (See Section~\ref{sec:rotation}).
Despite neglecting the important effect of radiative damping, the trend for the relative radial displacement at the surface do show some interesting correlations with the observed trend in SLF variability, though they generally proceed the wrong way, with increasing amplitude towards higher luminosities where we expect IGW to show the smallest effects (left panel in Fig.~\ref{fig:nuchar_core}).  

Next, we focus on the characteristic timescale of waves excited by core convection. 
We expect the maximum of the IGW flux to be launched at frequencies close to $\nuchar = (2\pi\tau_{\rm c})^{-1}$, with $\tau_{\rm c}$ defined in eqn.~\ref{eq:turnover}.
Fig.~\ref{fig:nuchar_core} shows that the predicted values of $\nuchar$ are in the range $0.02\dots0.008\,{\rm  d}^{-1}$ ($0.2\dots0.08\, \muhz$), in agreement with the results of \citet{Shiode:2013}.
These values are about 2 orders of magnitude smaller than the typical characteristic frequencies of  SLF variability observed by \citep{Bowman:2019,2019NatAs...3..760B}.

As mentioned earlier, a significant caveat in correlating core and surface quantities is that wave propagation through the stellar envelope affects the spectrum, changing the frequency of maximum power of waves reaching the surface \citep{Lecoanet:2020}.
Radiative damping efficiently suppresses low-frequency IGWs (because these waves have high radial wavenumbers), and the peak of the spectrum observed at the surface is expected to move to higher frequencies. The amplitude of this effect depends on the envelope properties, which are a function of $\logteff$ and $\logl$. Nevertheless we will assume that trends of $\nuchar$, as function of $\logteff$ and $\logl$, are still set by the core convective properties. 
Under this assumption, Fig.~\ref{fig:nuchar_core} shows that $\nuchar$ should increase with both $\logteff$ and $\logl$, so that the characteristic frequencies of peak IGWs should increase as stars evolve on the main sequence. This is exactly the opposite of what is observed:  the characteristic frequencies of SLF variability is largest for stars early on on the main sequence, and seem to decreases as stars evolve.  
Therefore either radiative damping is able to revert this trend or else the observed variability is unlikely to be caused by IGWs launched by the core.

\section{Macroturbulence in Magnetic Stars}

The hypothesis that subsurface convection is responsible for surface turbulence is corroborated by the match between trends in the FeCZ properties (i.e. convective velocities and turbulent pressure) and the observed micro and macroturbulence \citep{Cantiello:2009,Grassitelli:2015}.
One important test to this hypothesis is provided by the disappearance of macroturbulence in stars with surface magnetic fields above a critical strength \citep{10.1093/mnras/stt921}, closely corresponding to 
the critical field needed to stabilize the FeCZ~\citep{2019MNRAS.487.3904M,Jermyn:2020}. Note that in OB stars, the FeCZ is deeper and more vigorous than the H and He convection zones, so a magnetic field stabilizing the FeCZ will necessarily also stabilize the other subsurface convection zones.  

We can analogously define a critical magnetic field strength $B_{\rm crit}$ which suffices to reflect IGWs before they reach the photosphere. 
Using the dispersion relation for IGWs in a magnetic medium, the radial component of this field is~\citep{2015Sci...350..423F}
\begin{align}
	B_{r, \rm crit} = \frac{\omega}{2 k_r}\sqrt{4\pi \rho},
\end{align}
where $\omega=2\pi\nu$ is the angular frequency of the waves, $\rho$ is the density, and $k_r$ is the radial wave-number.
This critical field may be thought of as the field strength at which the Alfv{\'e}n frequency computed with the length-scale $1/k_r$ is comparable to the wave frequency and is therefore analogous to the effect of rotation, which enters in when the rotation angular velocity is faster than the wave frequency.
For IGWs the radial wave-number is related to the spherical harmonic degree $\ell$ by
\begin{align}\label{eq:kr}
	k_r \approx \frac{\sqrt{\ell(\ell+1)}}{r}\left(\frac{N}{\omega}\right),
\end{align}
where $N$ is the \brvs\ frequency.
So
\begin{align}
	\label{eq:Brcrit}
	B_{r, \rm crit} &\approx \frac{\omega^2 r}{N}\sqrt{\frac{4\pi \rho}{2\ell(\ell+1)}} \\
	&\approx \frac{\omega^2 r}{\ell N}\sqrt{4\pi \rho}.
\end{align}
Note that because this decreases with increasing $\ell$, all waves of a given frequency are reflected if the $\ell=1$ waves are reflected.

\begin{figure}[ht!]
\begin{center}
\includegraphics[width=1.0\columnwidth]{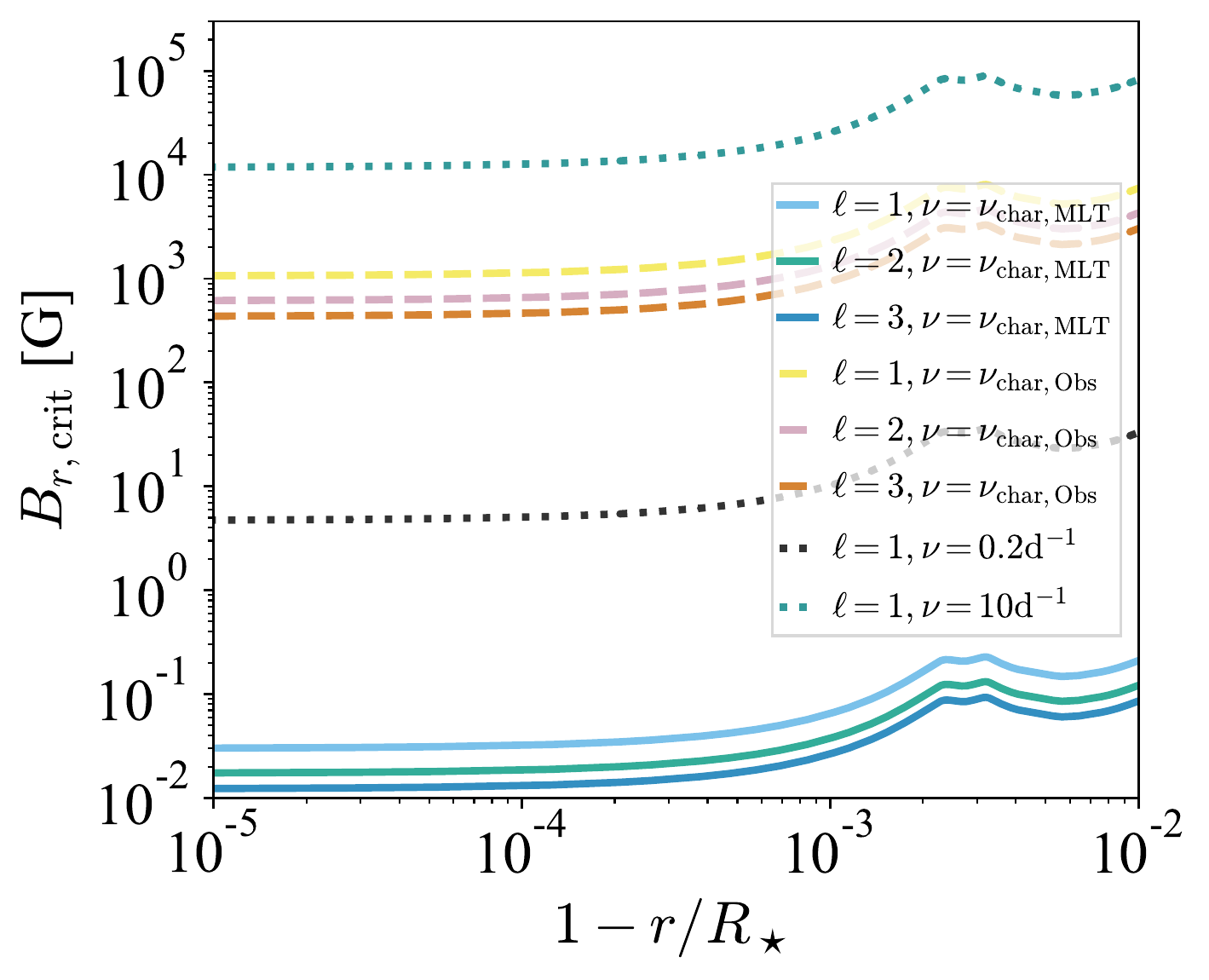}
\caption{\label{fig:magnetic} The critical radial magnetic field strength needed to reflect internal gravity waves of different $\ell$ and $\nu$ is show as a function of fractional depth for a $30 M_\odot$ stellar model at an age of $4\,\mathrm{Myr}$. The frequency $\nu_{\rm char,MLT}$ is that given by eqn.~\ref{eq:nuchar} but evaluated for the core convection zone instead of the subsurface FeCZ. The frequency $\nu_{\rm char,Obs}$ is a typical value of $3\,\mathrm{d}^{-1}$ for the observations. We further show frequencies of $0.2\,\mathrm{d^{-1}}$ and $10\,\mathrm{d^{-1}}$ as these span the full range of observed characteristic frequencies.}
\end{center}
\end{figure}

We computed $B_{r, \rm crit}$ for several values of $\ell$ and $\nu$ for a main-sequence model of a $30 M_\odot$ star as a function of radius, shown in Fig.~\ref{fig:magnetic}.
For frequencies comparable to those of core convection the critical magnetic field strength is of order $10^{-2}\,\mathrm{G}$ to $10^{-1}\,\mathrm{G}$.
Strong macroturbulence is observed in similar-mass O-type stars with magnetic fields up to $2.5\,\mathrm{kG}$ \citep[e.g. HD~191612,][]{10.1093/mnras/stt921}, so macroturbulence in those stars is unlikely to be due to IGWs coming from their cores if those waves have similar frequencies to that of core convection.

For frequencies comparable to the observed $\nu_{\rm char}$, on the order of $3\,\mathrm{d}^{-1}$, the critical magnetic field is much larger, on the order of $300\,\mathrm{G}$ to $1\,\mathrm{kG}$.
This makes an explanation of macroturbulence based on IGWs marginally inconsistent with observations showing strong macroturbulence up to field strengths of $2.5\,\mathrm{kG}$.
However, because eqn.~\ref{eq:Brcrit} is a strong function of $\omega$, and hence of $\nu$, it is possible that these strongly magnetized stars just have larger $\nu_{\rm char}$.
The full range of observed characteristic frequencies spans $0.2-10\,\mathrm{d^{-1}}$, corresponding to critical field strengths at the surface of $3\,\mathrm{G}-10\mathrm{kG}$ for the $\ell=1$ mode.
Thus while IGWs with lower frequencies are inconsistent with observations of strongly-magnetized stars with substantial macroturbulence, those at higher frequencies likely make it to the surface and could contribute to the observed macroturbulence.

A further prediction of this calculation is that, if IGWs are the cause of macroturbulence in these stars, we should expect the strength of macroturbulence to decline with increasing magnetic field strength as more and more modes are reflected before they reach the surface.
\citet{10.1093/mnras/stt921} find no such trend, though they do that macroturbulence vanishes when the magnetic field exceeds the FeCZ shutoff strength~\citep{2019MNRAS.487.3904M}, and this is consistent with observations of other strongly-magnetized O/B stars such as HD~215441~\citep{1989ApJ...344..876L} and HD~54879~\citep{2015A&A...581A..81C}, both of which show little or no macroturbulence and magnetic fields stronger than the theoretical shutoff field strength.
This again points against an explanation of macroturbulence based on core-generated IGWs.

\section{Rotation}\label{sec:rotation}

\subsection{Subsurface Convection}
The sample of stars discussed here have projected rotational velocities $v_{\rm eq} \sin i$ in the range 7 to 320 $\kms$, suggesting rotation could have an impact on the properties of subsurface convection.
The effect of rotation is usually measured by the convective Rossby number $R_0 = 1 / ( 2 \, \Omega \, \tau_c)$, where $\Omega$ is the stellar rotational frequency. For  $R_0 > 1$ we expect rotation to have a moderate to negligible impact. 
On the other hand when the rotational period becomes comparable or shorter than the convective turnover time ($R_0 \le 1$), the properties of convection can be altered substantially \citep[e.g.][]{Stevenson:1979,Augustson:2019}.   
The typical rotational period of OB stars is about 3 days (assuming a typical equatorial rotational velocity $v_{\rm eq} \approx 150 \kms$), while the convective turnover timescale in the FeCZ is a few hours (See e.g. Fig.~\ref{fig:nuchar_fecz}). 
Therefore the $R_0$ in the observed sample is likely in the range $1\dots10$. For these values, the convective velocities as calculated in the 1D MLT approximation are affected only at the $\sim 10\%$ level \citep[see e.g. Fig.4 in][]{Cantiello:2009}.
This said, the latitudinal structure of the FeCZ zone is substantially altered at the highest rotation rates \citep{Maeder:2008}, which could have an impact on the way these regions affect the stellar surface. 

\subsection{Core Convection}

In the stellar cores of intermediate and massive stars $R_0$ is very likely $<1$, so 
 rotation is expected to change the properties of convection \citep[e.g.][]{Stevenson:1979,Augustson:2019}.
At the same time, in the presence of rotation, gravity waves can be perturbed by the Coriolis acceleration and combine with inertial waves (gravito-inertial waves, GIWs). 
The stochastic excitation of gravity and GIWs by rotating convective zones was studied by \citet{Mathis:2014} and \citet{Augustson:2020}. 
The main result is that rotation can enhance the  amplitude of stochastically excited waves \citep{Mathis:2014}. 
The work of \citet{Neiner:2020} shows that in some rapidly rotating stars, stochastically excited GIWs from the core could explain part of the observed low-frequency variability. 

However, we note that it is unlikely that core-generated GIWs are responsible for the ubiquitous SLF variability. This is because the visibility of GIWs depends on the inclination of the stellar rotation axis respect to the observer. 
Internal waves cannot propagate at the poles for $\omega < 2\Omega$, where $\Omega$ is the stellar rotational frequency  and $\omega$ is the wave frequency. The largest wave flux is expected at low latitudes,
with the degree of  equatorial confinement proportional to $\Omega$.  This means that the propagation domain of subinertial ($\omega < 2\Omega$) GIWs excludes the pole and it becomes increasingly concentrated toward the equator for faster rotation
rate \citep{Dintrans:2000,Prat:2016,Augustson:2020}. 
Assuming that the stars in \citet{Bowman:2020} have spin vectors randomly oriented, some should be observed nearly pole-on.
Then if the SLF variability was due to GIWs, these objects would show very little power at frequencies less than $2\Omega$. 
Such a sharp decline in variability at low
frequencies is not observed in any of the stars, suggesting GIWs are not the culprit \citep{Lecoanet:2020}. 
A detailed study of the surface amplitude of waves excited by core convection in rotating early-type stars is beyond the scope of this paper.

\section{Discussion}\label{sec:discussion}
The presence of a subsurface convection zone can result in variability of photospheric properties via a number of processes, including wave excitation \citep{Cantiello:2009,Grassitelli:2015} and magnetic buoyancy \citep{Cantiello:2011}. 
A linear perturbative analysis is limited, since the turbulent fluctuations in these convective regions can be large \citep{Grassitelli:2015}. 
Multi dimensional simulations including radiation have been performed in a restricted range of the parameter space, and show that the full turbulent manifestation of these convective regions extends up to the stellar surface \citep{2015ApJ...813...74J,Jiang:2017,2018Natur.561..498J,Schultz:2020}.
In their calculations of OB stars envelopes including the stellar photosphere,  \citet{2015ApJ...813...74J}  observe turbulent velocities reaching the isothermal sound speed ($\approx 50 \kms$) at the stellar surface, demonstrating that velocity fields of amplitude comparable to the observed macroturbulence are naturally explained by the presence of the FeCZ. 

 Therefore it could be that the observed SLF variability and macroturbulence simply represent the direct manifestation of turbulent, radiation-dominated convection at the stellar surface.  
The simplified MLT treatment in our one dimensional calculations is unable to capture the complex phenomenology of these layers, but the fact that it can reproduce both the  timescales and the trends in amplitude of the observed SLF variabilty is compelling. It calls for extending the radiation hydrodynamics simulations to cover the parameter space of the TESS observations, in order to unravel the precise mechanism connecting subsurface convection zones to the observed surface variability.

\subsection{Perturbation Lengthscale}
If the perturbation is due to stochastically excited modes driven by the FeCZ \citep{Grassitelli:2015}, then we expect the largest fluctuations to be produced by modes with $\ell \lesssim 20$ \citep[see e.g.][]{Godart:2017}. 
The situation is different if instead the perturbation is provided by running waves or by convective motions extending to the stellar surface  \citep{Cantiello:2009,2015ApJ...813...74J}.  
In this case a good proxy for the typical scale of surface perturbations is provided by the size of convective cells in the FeCZ. 
This in turn is quantified by the average pressure scale height in the subsurface convection zone. Note that the pressure scale height only decreases slightly moving from the FeCZ to the stellar surface. 
In general one expects that velocity perturbations induced by the FeCZ should have scales that are  comparable or slightly larger than the line forming region, so  macroturbulence can be explained via this mechanism.
Rotation could also be responsible for organizing the convective flow on slightly larger scales (see Sec.~\ref{sec:rotation}).
Since  convective turbulence  also results in smaller-scale motions, the FeCZ could also be responsible for the excitation of surface microturbulence \citep{Cantiello:2009}. 

\begin{figure}[ht!]
\begin{center}
\includegraphics[width=1.0\columnwidth]{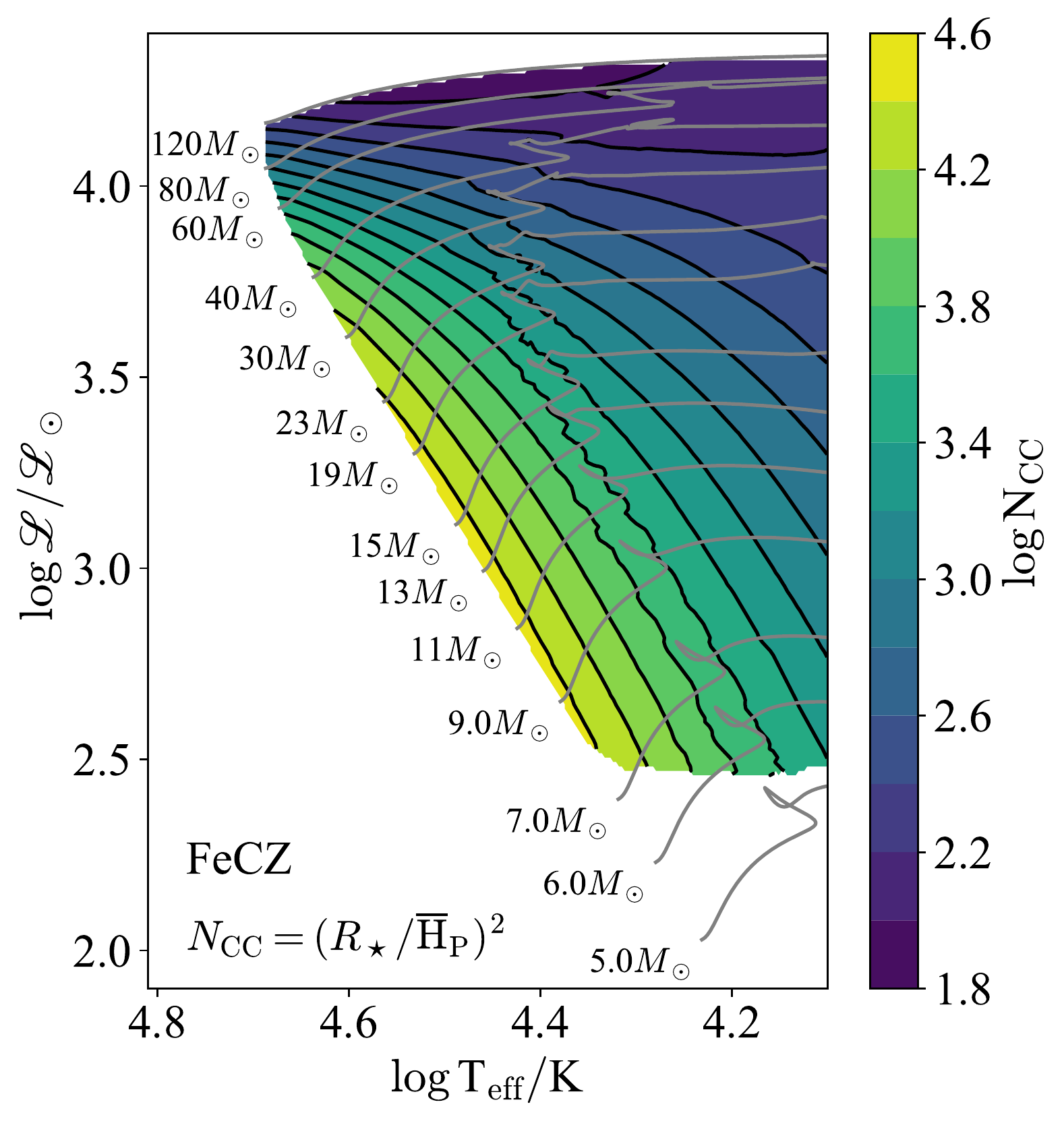}
\caption{\label{fig:cells} Number of convective cells in the FeCZ. This was calculated using the average pressure scale height in the FeCZ.}
\end{center}
\end{figure}

We show in Fig.~\ref{fig:cells} the number of convective cells $N_{\rm CC} = (R_\star/\overline{\rm H}_{\rm P})^2$,  calculated using the stellar radius  and the average pressure scale height in the FeCZ. 
We expect approximately $10^2...10^4$ convective cells in the FeCZ of OB stars.  Therefore the order of the perturbation $\ell \approx \sqrt{N_{\rm CC}}\approx 10...100$. 
While it might seem impossible for such high-degree perturbations to leave a visible signature on the stellar disc integrated properties, we point out that in this case the surface fluctuations are uncorrelated. So even high degree $\ell$ fluctuations do not  undergo the dramatic cancellation effects experienced by highly-correlated stellar oscillations. Similarly to granulation, we expect the amplitude of the integrated surface perturbations to scale as  $1/\sqrt{N_{\rm CC}}$.

It is important to distinguish the signal induced by  (sub)surface convection zones in OBA stars with the granulation pattern expected in cool stars with convective envelopes. The driving mechanism, properties, and location of these convection zones change substantially from late- to early-type stars  \citep{Cantiello:2019}. This is why is not surprising that the characteristic frequency of the SLF variability in the early-type stars observed by  \citet{Bowman:2019} and \citet{Szewczuk:2021}  does not follow the granulation scaling of \citet{Kjeldsen:1995}, which was derived for late-type stars. On the other hand we have shown here that the observed frequencies are consistent with the expectation of perturbations arising from subsurface convection zones (see Fig.~\ref{fig:nuchar_fecz}).

\subsection{Metallicity}
If the FeCZ is responsible for surface macroturbulence and SLF variability, then these phenomena should be affected by the stellar metallicity.  On the other hand we do not expect  a significant metallicity dependence in
the context of a core-convection origin.
The results of \citep{Bowman:2019} show that SLF variability is also observed in low-metallicity, LMC stars.  
The presence of the FeCZ depends on the luminosity and metallicity of the star, and in 1D stellar evolution calculations the FeCZ occurs above a luminosity of L$\approx10^{3.2} L_{\odot}$ for Z=0.02.
This corresponds to a zero age main sequence star of about 7$\mso$ (see e.g. Fig.~\ref{fig:macroturbulence}).
The LMC has a metallicity about half solar, and in models at Z=0.008 the FeCZ only appears at L$\approx10^{3.9} L_{\odot}$, corresponding to a zero age main sequence star of about 11$\mso$ \citep{Cantiello:2009}. In our models with metallicity Z=0.006 (See Appendix~\ref{appen:grids}) the FeCZ appears at  $\loglsun \approx 3.2$.

Note that these limits could move downward  in the presence of atomic diffusion and radiative acceleration \citep{Richer:2000}, or due to an upward revision of the uncertain values of Fe opacity \citep[e.g.][]{Bailey:2015}.  We notice that the TESS observations of LMC stars reported by \citet{2019NatAs...3..760B} reveal a trend of lower $\nuchar$ compared to the galactic sample (See their Fig.~4). Interestingly, this trend is reproduced by the characteristic frequency of convection in the FeCZ in our models (compare Fig.~\ref{fig:nuchar_fecz} with Fig.~\ref{fig:nuchar_fecz_LMC}). It will be interesting to see if the TESS observations  can probe the transition region between stars with and without a FeCZ, and determine a possible change in surface properties.
At the same time, X-shooter within the Ulysses program could more firmly establish the metallicity-dependence of the macroturbulent
line-broadening \citep{Penny:2009}.

It is important to note that some low-luminosity, main sequence A stars in \citet{Bowman:2019} show SLF variability. For these stars models do not predict the presence of the FeCZ. However, these stars still show (sub)surface convection zones triggered by ionization of H and He \citep{Cantiello:2019}.  The characteristic frequencies of convection in these regions are also in the range $\sim$tens of $\mu$Hz, although the  amplitude of the velocity fluctuations they can induce is much smaller than for the FeCZ, so it is not clear if they could be linked to observed SLF variability and macroturbulence. On the other hand, since the relative perturbation induced by subsurface convection is weakest in the regime  of A and late-B type \citep{Cantiello:2019,Jermyn:2020},
these stars are the best targets for detecting core-generated IGWs.  Depending on the amplitude of core generated IGWs, the impact of subsurface convection could well be subdominant in these objects, allowing for a detection. 

\subsection{Stochastic, Low-Frequency Variability in Evolved Stars}
SLF variability with similar properties to the massive main sequence stars discussed by \citep{Bowman:2019,Bowman:2020} was recently observed in evolved massive stars.
\citet{Naze:2021} detected SLF variability in both luminous blue variables (LBV) and Wolf-Rayet (WR) stars, and \citet{Dorn:2020} found the same photometric signature in yellow supergiants (YSG).  
Compared to OB stars, LBV, WR and YSG correspond to later stages of evolution. In particular, WR stars and YSG are likely burning helium in their cores. 

The internal structure of OB, LBV, WR and YSG stars changes dramatically,  and this can affect substantially the generation and propagation of internal gravity waves.
The amount of radiative damping is expected to change due to the large differences in envelope temperature and densities.  We recall here that the radiative damping rate $ \gamma_{\rm rad}$ for a traveling g-mode is given by
\begin{align*}\label{eqn:gammarad}
	&\rm \gamma_{rad} (\omega, \ell, r) = K_{\rm rad} (r) \, k_{r}^{2},\\ 
	&K_{\rm rad} (r) = \frac{16 \, \sigma \, T(r)^{3}}{3 \, \rho(r)^{2} \kappa(r) \, c_{p}(r)}.
\end{align*}
where $k_{r}$ is the radial wavenumber (eqn.~\ref{eq:kr}) and $\kappa$ is the opacity.
Compact WR stars have surface temperatures that can exceed $\approx 10^5$K, while OB stars and LBVs have effective temperatures $\approx 10^{4\dots4.7}$K, with LBVs found at the cooler end of this range.  The surface of extended, low-density, YSGs is cooler than $10^4$K.  With different types of core convection (H-burning vs He-burning) and radiative damping rates, it would be surprising if core-generated IGWs in e.g. OB and WR stars showed up at the surface with similar properties. On the other hand, the FeCZ is present below/at the surface in both OB and WR stars, and with similar velocities and convective turnover times. This seems to strengthen the main thesis of this work,  adding support to a (sub)surface origin of the observed SLF variability.

 The discussion is more complicated for LBVs and YSGs, since at lower temperatures other convective regions can become prominent \citep[driven by H and He recombination, see e.g.][]{2018Natur.561..498J}.  \citet{Dorn:2020} disfavor a near-surface origin for the observed SLF variability in YSG, on the ground that the observed timescale do not follow the predicted scaling for granulation \citep{Kallinger:2014}.  However, we believe that the scaling of \citet{Kallinger:2014} is not applicable in the regime explored by \citet{Dorn:2020}. This scaling was derived for a sample of red giants and solar-like stars, which all have temperatures well-below the temperature for the recombination of hydrogen, and it is not directly applicable to earlier
spectral subtypes. 

Interestingly, the characteristic timescales of 0.1-1 days found by \citet{Dorn:2020} at $\logteff \approx$4 are consistent with the convective turnover timescale in the FeCZ  (e.g. Fig.~\ref{fig:nuchar_fecz}). Moreover, in their lower temperature sample ($\logteff < 3.75$) the rapid increase in amplitudes and characteristic frequencies  of the variability  is in agreement with the development of near-surface convection induced by the large opacities associated with the recombination of hydrogen \citep{Grassitelli:2015b}. We suggest that surface and near-surface convection could be indeed responsible for the variability observed by \citet{Dorn:2020}, and we plan to systematically study the properties of (sub)surface convection in these evolved, cool stars in future work.

\subsection{Towards a unified model for surface phenomena in massive stars}
The presence of subsurface convection can simultaneously account for a large variety of puzzling phenomena that appear ubiquitous at the surface of early-type stars.
\begin{itemize}
\item {\bf Microturbulence and Macroturbulence} can be accounted for by velocity fields excited by the underlying FeCZ subsurface convection zone \citep{Cantiello:2009,Grassitelli:2015,2015ApJ...813...74J}. The only stars with no macroturbulence appear to have magnetic fields strong enough to shut off convection in the FeCZ, while stars with slightly lower surface magnetic fields show normal values of macroturbulence \citep{10.1093/mnras/stt921}. 
\item {\bf Bright spots}  in early-type stars have been observed  \citep[e.g.][]{Ramiaramanantsoa:2014}, and can be explained with the presence of magnetic spots rising from subsurface convective layers \citep{Cantiello:2011,Cantiello:2019}. 
\item {\bf Discrete absorption components (DACs)} in UV spectra \citep[e.g.][]{Howarth:1989,Cranmer:1996,Fullerton:1997,Kaper:1997} can then be caused by the aforementioned bright spots \citep{Cantiello:2011} and associated prominences \citep{Sudnik:2016}. 
\item {\bf Line profile variability}   is another ubiquitous phenomena in hot stars \citep{Fullerton:1996}, and can be explained by surface velocity and density perturbations seeded by subsurface convection \citep[e.g.][]{Cantiello:2009,2015ApJ...813...74J}.
\item {\bf Wind clumping} can also be seeded by these surface density and velocity perturbations, which are amplified by the development of instabilities in the stellar wind \citep[][but see also \citet{Sundqvist:2013}]{Owocki:1988}.
\item {\bf SLF variability}  can also be caused by the presence of  subsurface convection zones, as discussed in this work.  
\end{itemize}
An economical hypothesis emerges: the presence of subsurface convection, and in particular of the FeCZ, could be the common underlying physical cause for the appearance of turbulence, magnetic spots, SLF variability, as well spectroscopic variability in early-type stars.

\section{Conclusions}

We used one-dimensional, non-rotating stellar evolution calculations  to study the predicted trends in the properties of subsurface convection in the spectroscopic H-R diagram.
We found that the trends of relative convective flux and convective turnover timescale in the FeCZ of our models match very well the trends in timescale 
and amplitude of stochastic, low-frequency  photometric variability in OB stars observed by TESS and K2.

Similar to previous works, we  show that the observed trends in stellar macroturbulence are also well reproduced assuming the FeCZ is its driver. This connection is also supported by the observations of strongly magnetized early-type stars,
showing no macroturbulence only for magnetic fields above the critical value required to shut-off turbulent convection in the FeCZ. We find that IGWs coming from the stellar core would be reflected or damped for values of the magnetic field well below this critical value. The fact that stars with strong but subcritical magnetic fields show typical values of macroturbulence points against a convective core origin of this surface perturbation. 
In the presence of rotation, GIWs are also expected to propagate and reach the stellar surface. These waves are increasingly confined to stellar equatorial regions in rapidly-rotating stars, and for stars seen close to pole-on a  sharp decline in their variability is expected below twice their rotational frequency. This feature is not detected in any of the early-type stars observed, suggesting GIWs are also unlikely to explain the observed surface variability.

Overall the observations support a picture in which subsurface convection, and in particular the FeCZ, is responsible for the ubiquitous low-frequency, stochastic photometric variability and macroturbulence detected in OB stars. 
These surface manifestations join a number of phenomena observed in early-type stars and attributed to the presence of subsurface convection, including the observations of (magnetic) bright spots as well as wind and spectroscopic variabilty. Radiation (magneto)hydrodynamics simulations of the outer envelope regions of early-type stars are required to understand the details of how  subsurface convection zones cause the observed surface perturbations.

\acknowledgments
We thank the anonymous referee for a constructive report which helped improve the manuscript. We also thank Evan Anders for useful discussions on stellar convection.
The Center for Computational Astrophysics at the Flatiron Institute is supported by the Simons Foundation.
This research was supported in part by the National Science Foundation under Grant No. NSF PHY-1748958 and by the Gordon and Betty Moore Foundation through Grant GBMF7392. DL is supported by a Lyman Spitzer Jr.~Fellowship.

\appendix
\section{Software Details}
\label{sec:software}

Calculations were done with \code{MESA} version 15140.
The \code{MESA} EOS is a blend of the OPAL \citep{Rogers2002}, SCVH
\citep{Saumon1995}, FreeEOS \citep{Irwin2004}, HELM \citep{Timmes2000},
and PC \citep{Potekhin2010} EOSes.
Radiative opacities are primarily from OPAL \citep{Iglesias1993,
Iglesias1996}, with low-temperature data from \citet{Ferguson2005}
and the high-temperature, Compton-scattering dominated regime by
\citet{Buchler1976}.  Electron conduction opacities are from
\citet{Cassisi2007}.
Nuclear reaction rates are from JINA REACLIB \citep{Cyburt2010} plus
additional tabulated weak reaction rates \citet{Fuller1985, Oda1994,
Langanke2000}.  Screening is included via the prescription of \citet{Chugunov2007}.
Thermal neutrino loss rates are from \citet{Itoh1996}.
The inlists, processing scripts, and model output are available as a ZENODO repository: \dataset[https://zenodo.org/record/4676427]{\doi{10.5281/zenodo.4676427}}.

\section{Radial Displacement from IGWs}
\label{appen:xir}
Let's consider traveling internal gravity waves in the WKB approximation. 
The wave flux  is  $F_{\rm IGW} = \frac{1}{2} \rho \, |u|^2 u_{g,r}$
 with the group velocity $u_{g,r} = \partial \omega/\partial \kappa $.
Since  $u_{g,r} = (\omega/N)^2 \sqrt{N^2 - \omega^2} / k_h$,  $|u|^2 = (N/\omega)^2 u_r^2$, and 
$u_r = \omega \xi_r$ we obtain: 
\begin{equation}
F_{\rm IGW} = \frac{1}{2} \,\rho\, \xi_r^2 \omega^2 \frac{\sqrt{N^2-\omega^2}}{k_h}.
\end{equation}
Using  $k_h = \sqrt{\ell(\ell+1)}/r$ we can then write the relative radial displacement as:
\begin{equation}
\frac{\xi_r}{r} = \left[ \frac{F_{\rm IGW} \sqrt{\ell(\ell+1)}}{1/2 \rho \,r^3 \omega^2 \sqrt{N^2-\omega^2}} \right]^{1/2} \propto \sqrt{\frac{F_{\rm IGW}}{F_0}}
\end{equation}

\section{Grids at Different Metallicities}
\label{appen:grids}
\begin{figure*}[htp]
  \centering
    \subfloat{\includegraphics[width=0.495\columnwidth]{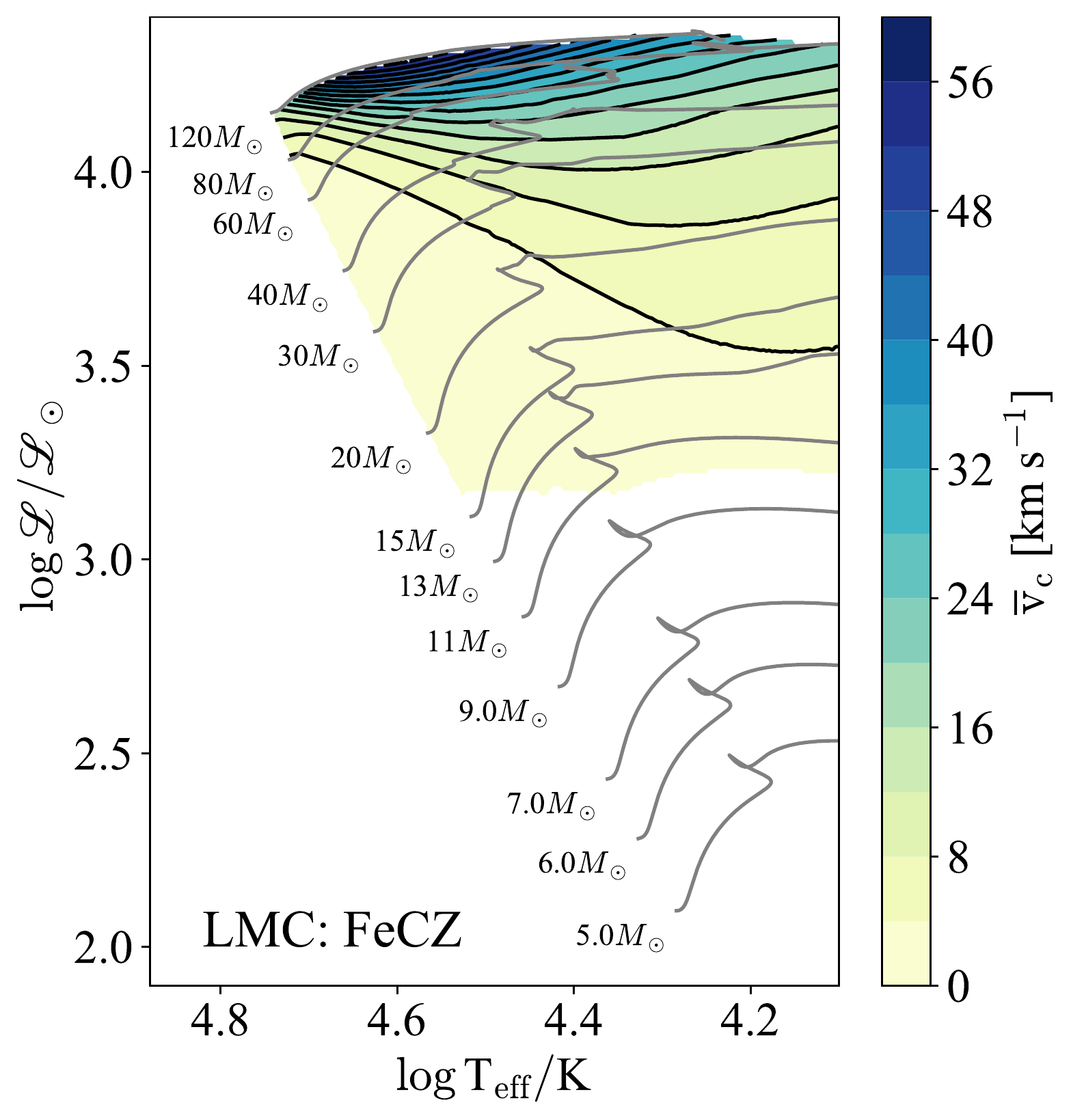}}\hfill
  \subfloat{\includegraphics[width=0.5\columnwidth]{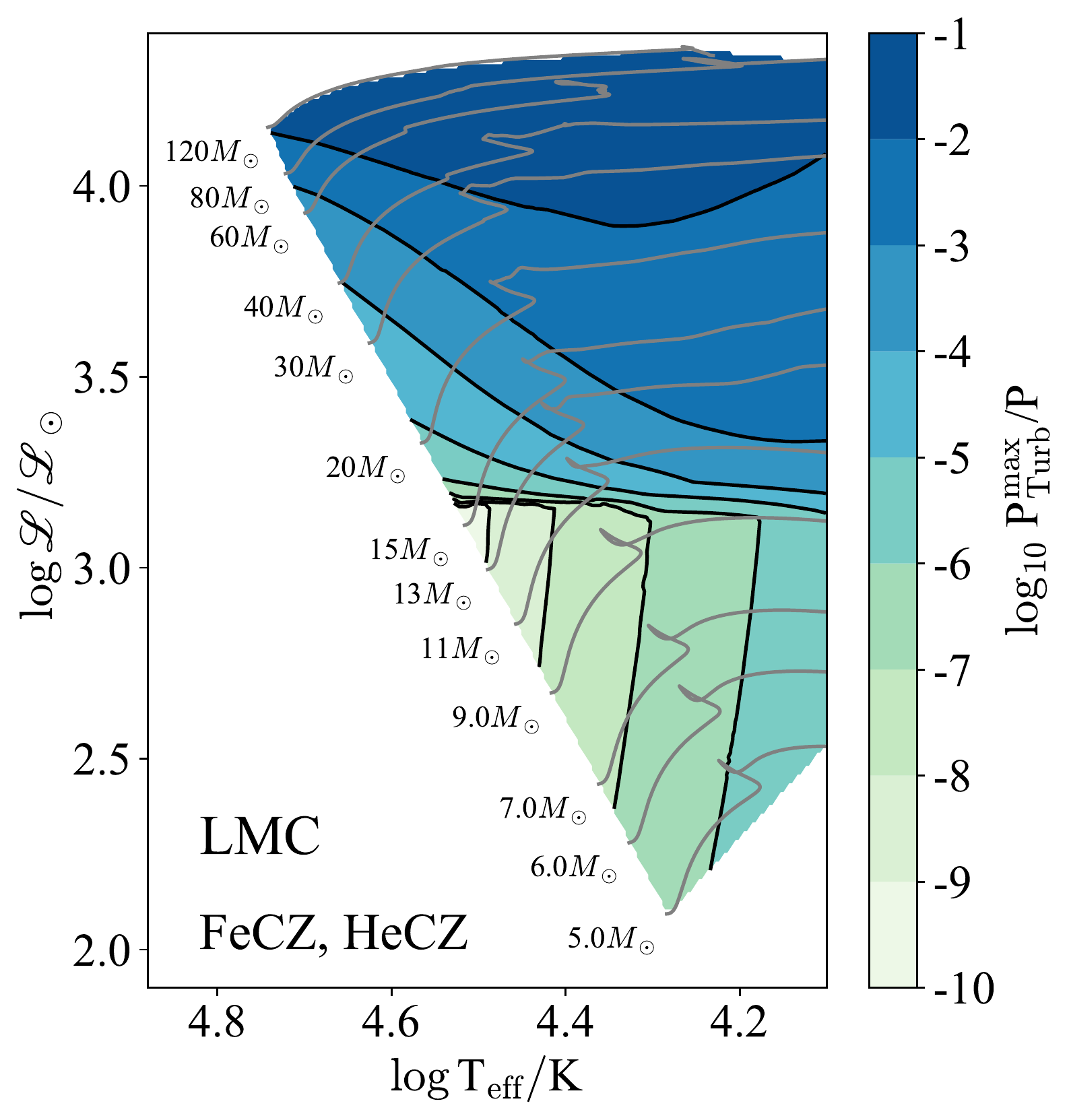}}
 \caption{\label{fig:vconv_LMC} Left: Average convective velocities in the FeCZ  as function of the location of stellar models in the spectroscopic H-R Diagram for models with Z=0.006 (LMC). The FeCZ is absent in models with $\loglsun \lesssim 3.2$. Right: Same as left, but showing predictions for the maximum of the ratio between turbulent pressure and total pressure in any subsurface convection zone. The FeCZ largely dominates, except for stars at  low luminosity where turbulent pressure is provided by a helium convection zone (HeCZ).} 
 \end{figure*}

\begin{figure*}[htp]
  \centering
    \subfloat{\includegraphics[width=0.5\columnwidth]{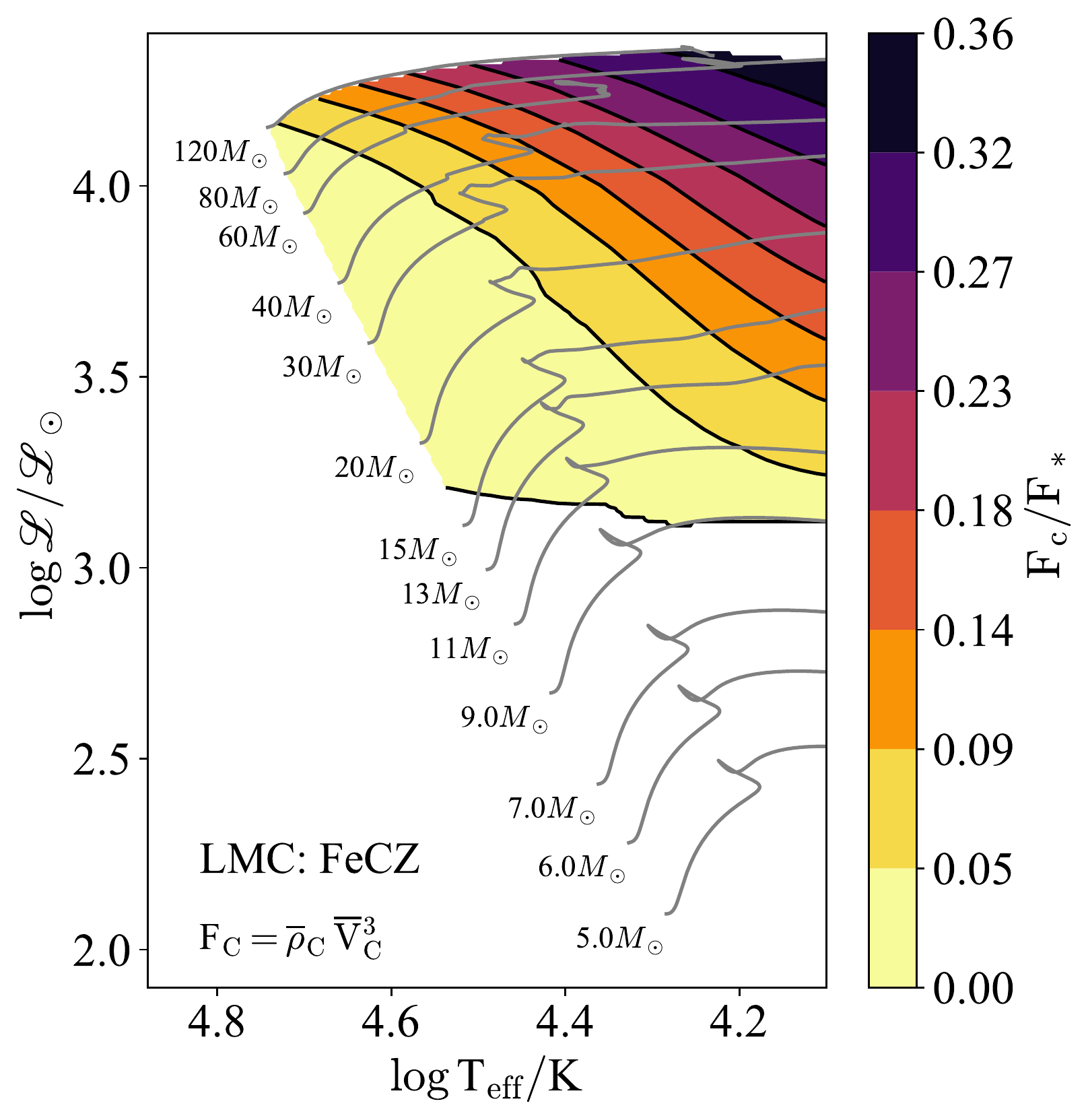}}\hfill
  \subfloat{\includegraphics[width=0.49\columnwidth]{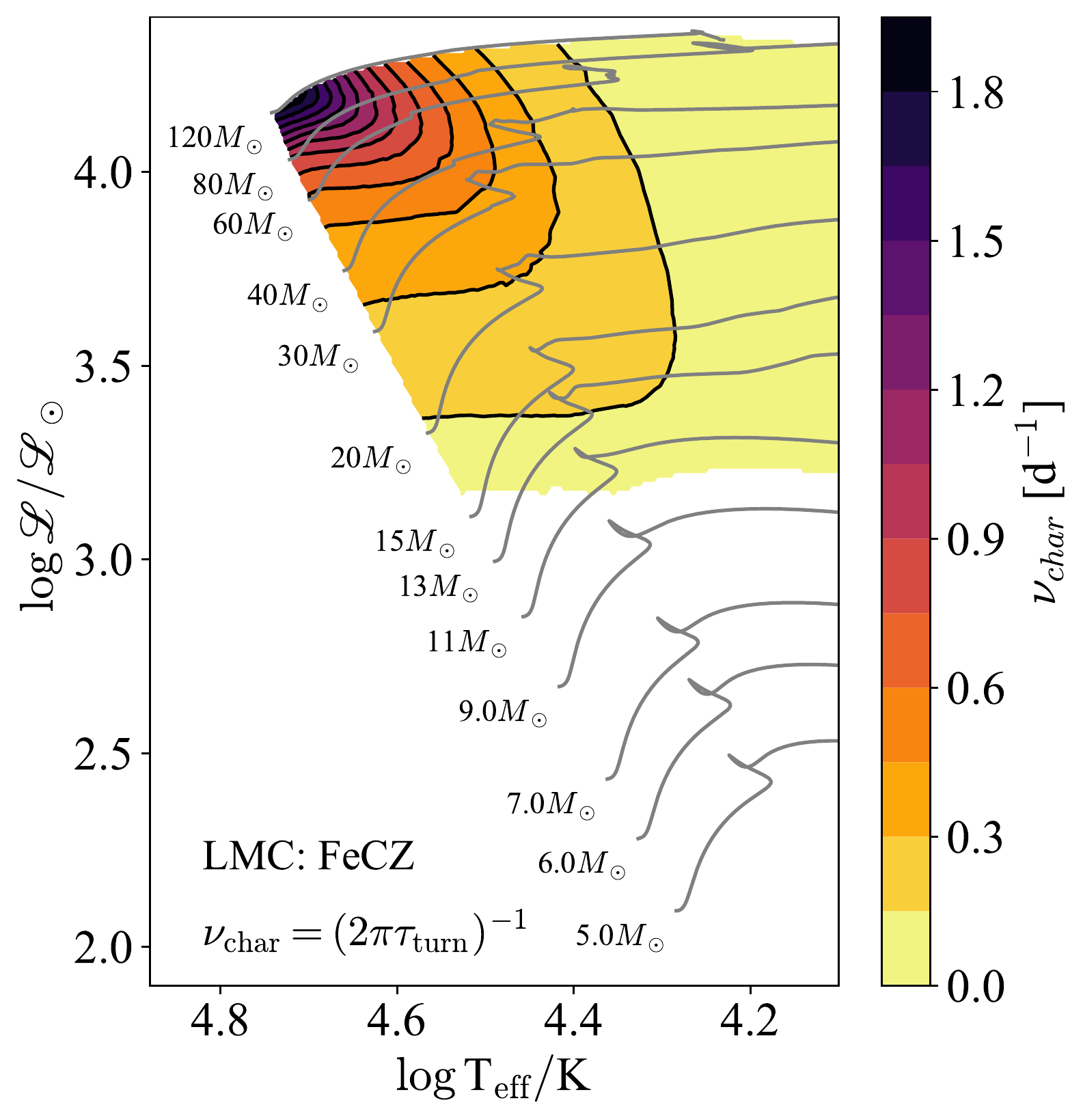}}
   \caption{  Left panel:  Ratio of FeCZ convective flux to the total stellar flux in the spectroscopic H-R Diagram  for models with Z=0.006 (LMC). Right panel: Characteristic frequency $\nu_{\rm char}$ in the FeCZ as a function of the location of stellar models in the spectroscopic H-R Diagram (black contour lines). Evolutionary tracks are shown as grey solid lines.}
    \label{fig:nuchar_fecz_LMC}
\end{figure*}

\begin{figure*}[htp]
  \centering
    \subfloat{\includegraphics[width=0.48\columnwidth]{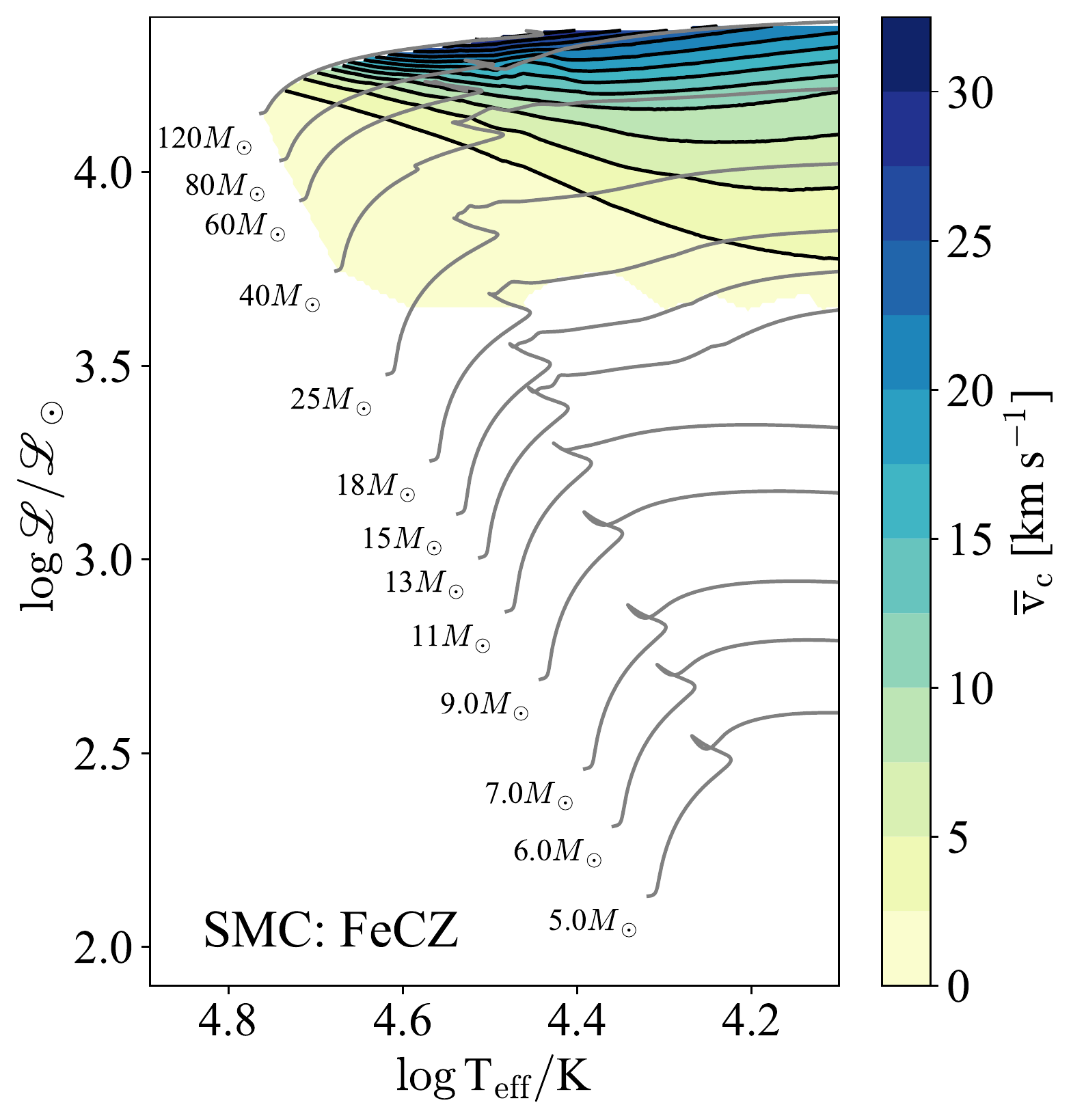}}\hfill
  \subfloat{\includegraphics[width=0.51\columnwidth]{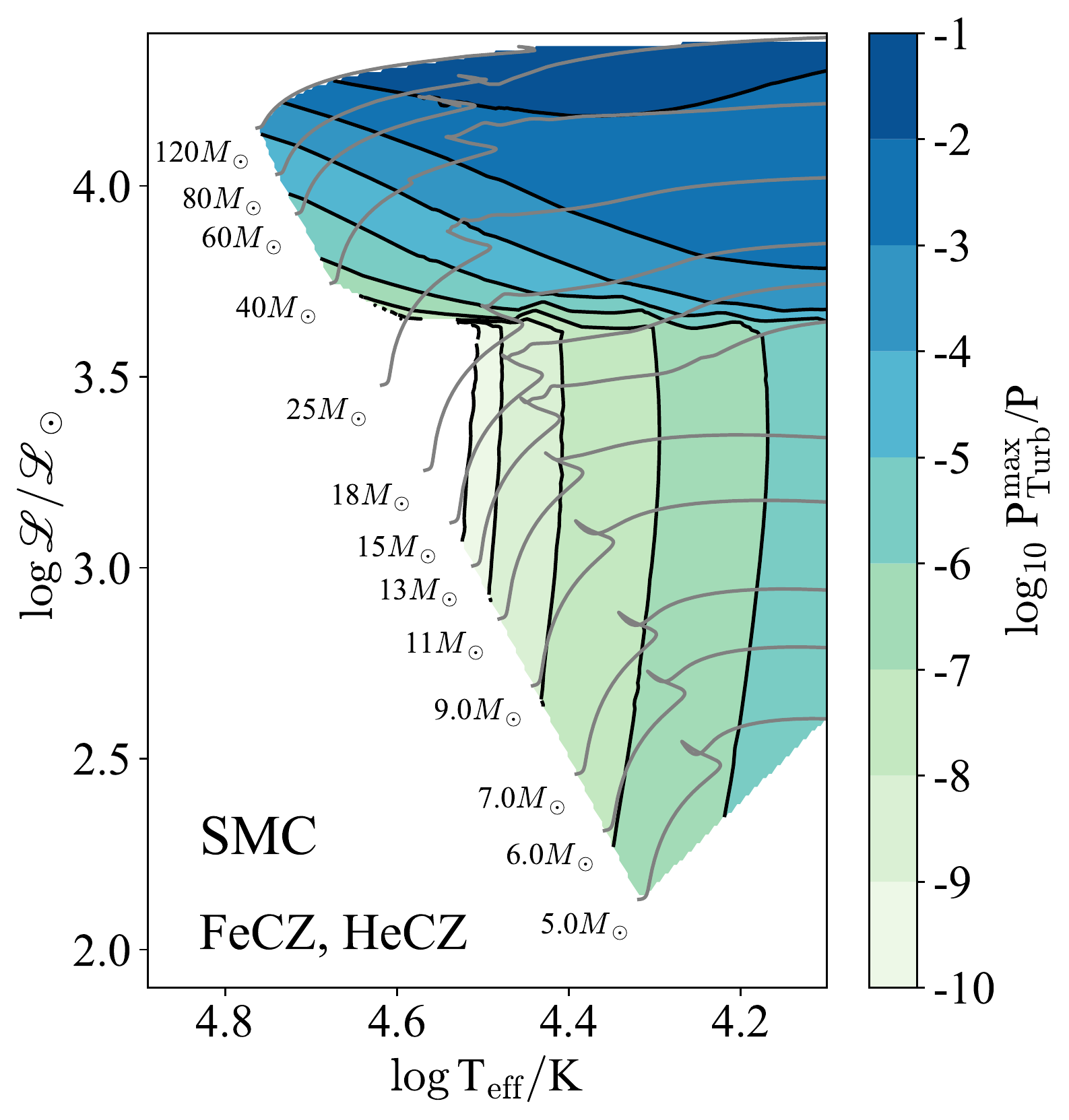}}
 \caption{\label{fig:vconv_SMC} Left: Average convective velocities in the FeCZ  as function of the location of stellar models in the spectroscopic H-R Diagram for models with Z=0.002 (SMC). The FeCZ is absent in models with $\loglsun \lesssim 3.2$. Right: Same as left, but showing predictions for the maximum of the ratio between turbulent pressure and total pressure in any subsurface convection zone. The FeCZ largely dominates, except for stars at  low luminosity where turbulent pressure is provided by a helium convection zone.} 
 \end{figure*}

\begin{figure*}[htp]
  \centering
    \subfloat{\includegraphics[width=0.5\columnwidth]{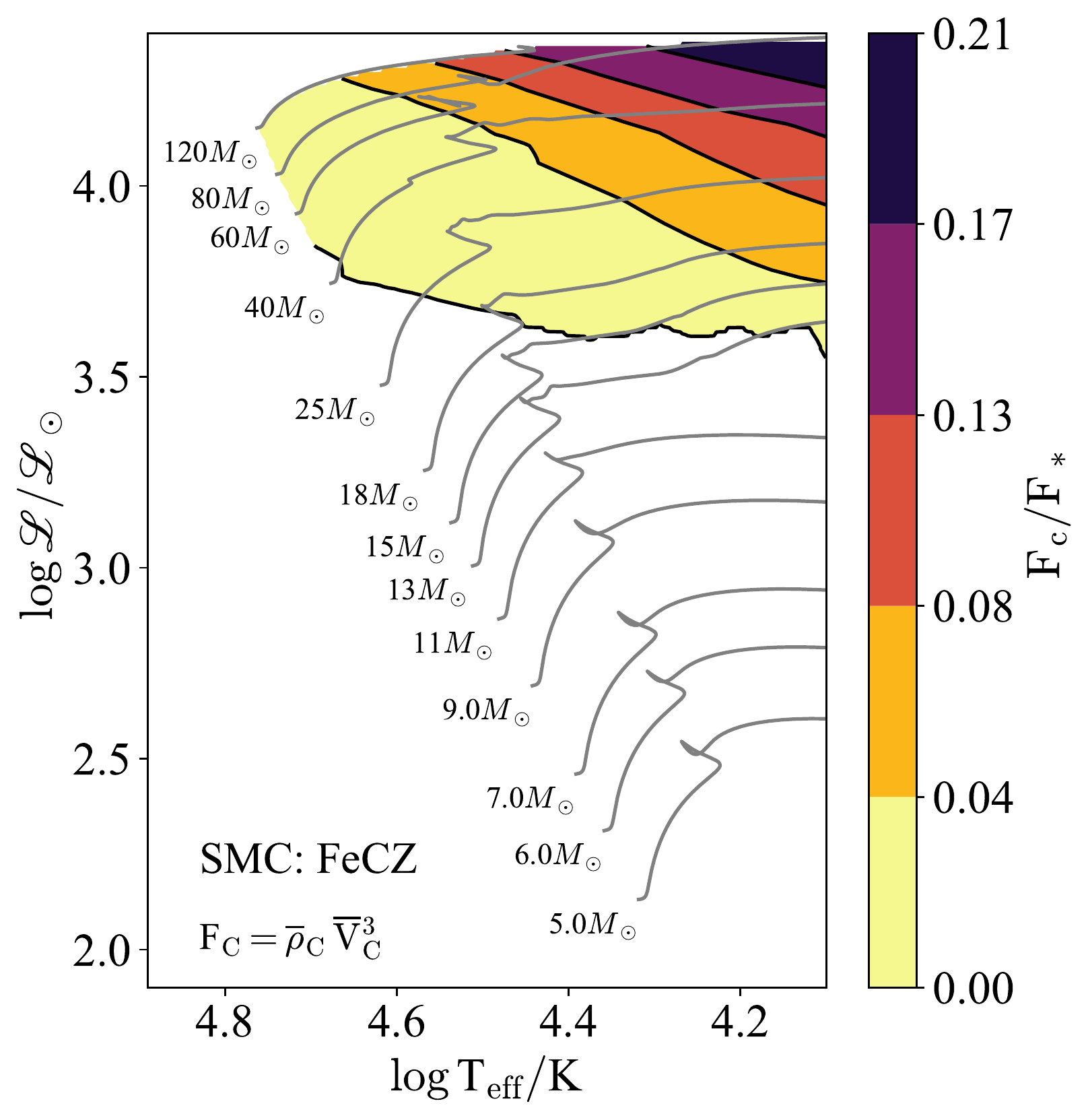}}\hfill
  \subfloat{\includegraphics[width=0.49\columnwidth]{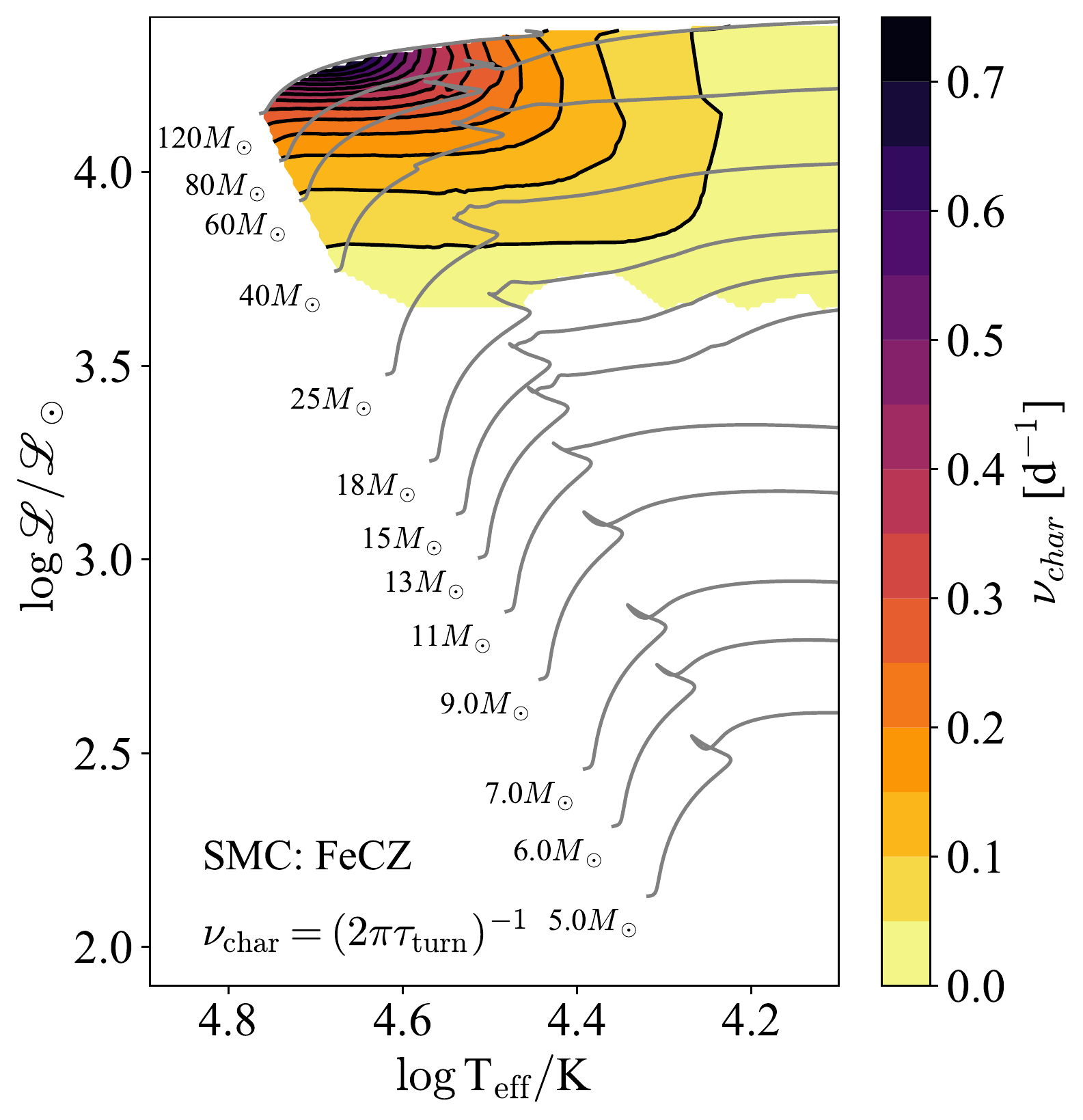}}
   \caption{  Left panel:  Ratio of FeCZ convective flux to the total stellar flux in the spectroscopic H-R Diagram  for models with Z=0.002 (SMC). Right panel: Characteristic frequency $\nu_{\rm char}$ in the FeCZ as a function of the location of stellar models in the spectroscopic H-R Diagram (black contour lines). Evolutionary tracks are shown as grey solid lines.}
    \label{fig:nuchar_fecz_SMC}
\end{figure*}

Here we present results for models at metallicities of Z=0.006 and Z=0.002,
representing early-type stars in the Large Magellanic Cloud (LMC, Fig.~\ref{fig:vconv_LMC} and \ref{fig:nuchar_fecz_LMC}) and Small Magellanic Cloud (SMC, Fig.~\ref{fig:vconv_SMC} and \ref{fig:nuchar_fecz_SMC}) respectively  \citep[e.g.][]{2013MNRAS.433.1114Y}. The initial helium content of the grids is
Y=0.2559 (LMC) and Y=0.2508 (SMC). The metallicity is initialized scaling the standard solar composition of \citet{1998SSRv...85..161G}.

\bibliographystyle{aasjournal}
\bibliography{refs.bib}

\begin{thebibliography}{}
\expandafter\ifx\csname natexlab\endcsname\relax\def\natexlab#1{#1}\fi
\providecommand{\url}[1]{\href{#1}{#1}}
\providecommand{\dodoi}[1]{doi:~\href{http://doi.org/#1}{\nolinkurl{#1}}}
\providecommand{\doeprint}[1]{\href{http://ascl.net/#1}{\nolinkurl{http://ascl.net/#1}}}
\providecommand{\doarXiv}[1]{\href{https://arxiv.org/abs/#1}{\nolinkurl{https://arxiv.org/abs/#1}}}

\bibitem[{{Abbott} {et~al.}(2016){Abbott}, {Abbott}, {Abbott}, {Abernathy},
  {Acernese}, {Ackley}, {Adams}, {Adams}, {Addesso}, {Adhikari}, {Adya},
  {Affeldt}, {Agathos}, {Agatsuma}, {Aggarwal}, {Aguiar}, {Aiello}, {Ain},
  {Ajith}, {Allen}, {Allocca}, {Altin}, {Anderson}, {Anderson}, {Arai},
  {Arain}, {Araya}, {Arceneaux}, {Areeda}, {Arnaud}, {Arun}, {Ascenzi},
  {Ashton}, {Ast}, {Aston}, {Astone}, {Aufmuth}, {Aulbert}, {Babak}, {Bacon},
  {Bader}, {Baker}, {Baldaccini}, {Ballardin}, {Ballmer}, {Barayoga},
  {Barclay}, {Barish}, {Barker}, {Barone}, {Barr}, {Barsotti}, {Barsuglia},
  {Barta}, {Bartlett}, {Barton}, {Bartos}, {Bassiri}, {Basti}, {Batch},
  {Baune}, {Bavigadda}, {Bazzan}, {Behnke}, {Bejger}, {Belczynski}, {Bell},
  {Bell}, {Berger}, {Bergman}, {Bergmann}, {Berry}, {Bersanetti}, {Bertolini},
  {Betzwieser}, {Bhagwat}, {Bhandare}, {Bilenko}, {Billingsley}, {Birch},
  {Birney}, {Birnholtz}, {Biscans}, {Bisht}, {Bitossi}, {Biwer}, {Bizouard},
  {Blackburn}, {Blair}, {Blair}, {Blair}, {Bloemen}, {Bock}, {Bodiya}, {Boer},
  {Bogaert}, {Bogan}, {Bohe}, {Bojtos}, {Bond}, {Bondu}, {Bonnand}, {Boom},
  {Bork}, {Boschi}, {Bose}, {Bouffanais}, {Bozzi}, {Bradaschia}, {Brady},
  {Braginsky}, {Branchesi}, {Brau}, {Briant}, {Brillet}, {Brinkmann},
  {Brisson}, {Brockill}, {Brooks}, {Brown}, {Brown}, {Brown}, {Buchanan},
  {Buikema}, {Bulik}, {Bulten}, {Buonanno}, {Buskulic}, {Buy}, {Byer},
  {Cabero}, {Cadonati}, {Cagnoli}, {Cahillane}, {Bustillo}, {Callister},
  {Calloni}, {Camp}, {Cannon}, {Cao}, {Capano}, {Capocasa}, {Carbognani},
  {Caride}, {Casanueva Diaz}, {Casentini}, {Caudill}, {Cavagli{\`a}},
  {Cavalier}, {Cavalieri}, {Cella}, {Cepeda}, {Baiardi}, {Cerretani},
  {Cesarini}, {Chakraborty}, {Chalermsongsak}, {Chamberlin}, {Chan}, {Chao},
  {Charlton}, {Chassande-Mottin}, {Chen}, {Chen}, {Cheng}, {Chincarini},
  {Chiummo}, {Cho}, {Cho}, {Chow}, {Christensen}, {Chu}, {Chua}, {Chung},
  {Ciani}, {Clara}, {Clark}, {Cleva}, {Coccia}, {Cohadon}, {Colla}, {Collette},
  {Cominsky}, {Constancio}, {Conte}, {Conti}, {Cook}, {Corbitt}, {Cornish},
  {Corsi}, {Cortese}, {Costa}, {Coughlin}, {Coughlin}, {Coulon}, {Countryman},
  {Couvares}, {Cowan}, {Coward}, {Cowart}, {Coyne}, {Coyne}, {Craig},
  {Creighton}, {Creighton}, {Cripe}, {Crowder}, {Cruise}, {Cumming},
  {Cunningham}, {Cuoco}, {Dal Canton}, {Danilishin}, {D'Antonio}, {Danzmann},
  {Darman}, {Da Silva Costa}, {Dattilo}, {Dave}, {Daveloza}, {Davier},
  {Davies}, {Daw}, {Day}, {De}, {DeBra}, {Debreczeni}, {Degallaix}, {De
  Laurentis}, {Del{\'e}glise}, {Del Pozzo}, {Denker}, {Dent}, {Dereli},
  {Dergachev}, {DeRosa}, {De Rosa}, {DeSalvo}, {Dhurandhar}, {D{\'\i}az}, {Di
  Fiore}, {Di Giovanni}, {Di Lieto}, {Di Pace}, {Di Palma}, {Di Virgilio},
  {Dojcinoski}, {Dolique}, {Donovan}, {Dooley}, {Doravari}, {Douglas},
  {Downes}, {Drago}, {Drever}, {Driggers}, {Du}, {Ducrot}, {Dwyer}, {Edo},
  {Edwards}, {Effler}, {Eggenstein}, {Ehrens}, {Eichholz}, {Eikenberry},
  {Engels}, {Essick}, {Etzel}, {Evans}, {Evans}, {Everett}, {Factourovich},
  {Fafone}, {Fair}, {Fairhurst}, {Fan}, {Fang}, {Farinon}, {Farr}, {Farr},
  {Favata}, {Fays}, {Fehrmann}, {Fejer}, {Feldbaum}, {Ferrante}, {Ferreira},
  {Ferrini}, {Fidecaro}, {Finn}, {Fiori}, {Fiorucci}, {Fisher}, {Flaminio},
  {Fletcher}, {Fong}, {Fournier}, {Franco}, {Frasca}, {Frasconi}, {Frede},
  {Frei}, {Freise}, {Frey}, {Frey}, {Fricke}, {Fritschel}, {Frolov}, {Fulda},
  {Fyffe}, {Gabbard}, {Gair}, {Gammaitoni}, {Gaonkar}, {Garufi}, {Gatto},
  {Gaur}, {Gehrels}, {Gemme}, {Gendre}, {Genin}, {Gennai}, {George}, {Gergely},
  {Germain}, {Ghosh}, {Ghosh}, {Ghosh}, {Giaime}, {Giardina}, {Giazotto},
  {Gill}, {Glaefke}, {Gleason}, {Goetz}, {Goetz}, {Gondan}, {Gonz{\'a}lez},
  {Castro}, {Gopakumar}, {Gordon}, {Gorodetsky}, {Gossan}, {Gosselin},
  {Gouaty}, {Graef}, {Graff}, {Granata}, {Grant}, {Gras}, {Gray}, {Greco},
  {Green}, {Greenhalgh}, {Groot}, {Grote}, {Grunewald}, {Guidi}, {Guo},
  {Gupta}, {Gupta}, {Gushwa}, {Gustafson}, {Gustafson}, {Hacker}, {Hall},
  {Hall}, {Hammond}, {Haney}, {Hanke}, {Hanks}, {Hanna}, {Hannam}, {Hanson},
  {Hardwick}, {Harms}, {Harry}, {Harry}, {Hart}, {Hartman}, {Haster},
  {Haughian}, {Healy}, {Heefner}, {Heidmann}, {Heintze}, {Heinzel}, {Heitmann},
  {Hello}, {Hemming}, {Hendry}, {Heng}, {Hennig}, {Heptonstall}, {Heurs},
  {Hild}, {Hoak}, {Hodge}, {Hofman}, {Hollitt}, {Holt}, {Holz}, {Hopkins},
  {Hosken}, {Hough}, {Houston}, {Howell}, {Hu}, {Huang}, {Huerta}, {Huet},
  {Hughey}, {Husa}, {Huttner}, {Huynh-Dinh}, {Idrisy}, {Indik}, {Ingram},
  {Inta}, {Isa}, {Isac}, {Isi}, {Islas}, {Isogai}, {Iyer}, {Izumi}, {Jacobson},
  {Jacqmin}, {Jang}, {Jani}, {Jaranowski}, {Jawahar}, {Jim{\'e}nez-Forteza},
  {Johnson}, {Johnson-McDaniel}, {Jones}, {Jones}, {Jonker}, {Ju}, {Haris},
  {Kalaghatgi}, {Kalogera}, {Kandhasamy}, {Kang}, {Kanner}, {Karki},
  {Kasprzack}, {Katsavounidis}, {Katzman}, {Kaufer}, {Kaur}, {Kawabe},
  {Kawazoe}, {K{\'e}f{\'e}lian}, {Kehl}, {Keitel}, {Kelley}, {Kells},
  {Kennedy}, {Keppel}, {Key}, {Khalaidovski}, {Khalili}, {Khan}, {Khan},
  {Khan}, {Khazanov}, {Kijbunchoo}, {Kim}, {Kim}, {Kim}, {Kim}, {Kim}, {Kim},
  {King}, {King}, {Kinzel}, {Kissel}, {Kleybolte}, {Klimenko}, {Koehlenbeck},
  {Kokeyama}, {Koley}, {Kondrashov}, {Kontos}, {Koranda}, {Korobko}, {Korth},
  {Kowalska}, {Kozak}, {Kringel}, {Krishnan}, {Kr{\'o}lak}, {Krueger}, {Kuehn},
  {Kumar}, {Kumar}, {Kuo}, {Kutynia}, {Kwee}, {Lackey}, {Landry}, {Lange},
  {Lantz}, {Lasky}, {Lazzarini}, {Lazzaro}, {Leaci}, {Leavey}, {Lebigot},
  {Lee}, {Lee}, {Lee}, {Lee}, {Lenon}, {Leonardi}, {Leong}, {Leroy},
  {Letendre}, {Levin}, {Levine}, {Li}, {Libson}, {Littenberg}, {Lockerbie},
  {Logue}, {Lombardi}, {London}, {Lord}, {Lorenzini}, {Loriette}, {Lormand},
  {Losurdo}, {Lough}, {Lousto}, {Lovelace}, {L{\"u}ck}, {Lundgren}, {Luo},
  {Lynch}, {Ma}, {MacDonald}, {Machenschalk}, {MacInnis}, {Macleod},
  {Maga{\~n}a-Sandoval}, {Magee}, {Mageswaran}, {Majorana}, {Maksimovic},
  {Malvezzi}, {Man}, {Mandel}, {Mandic}, {Mangano}, {Mansell}, {Manske},
  {Mantovani}, {Marchesoni}, {Marion}, {M{\'a}rka}, {M{\'a}rka}, {Markosyan},
  {Maros}, {Martelli}, {Martellini}, {Martin}, {Martin}, {Martynov}, {Marx},
  {Mason}, {Masserot}, {Massinger}, {Masso-Reid}, {Matichard}, {Matone},
  {Mavalvala}, {Mazumder}, {Mazzolo}, {McCarthy}, {McClelland}, {McCormick},
  {McGuire}, {McIntyre}, {McIver}, {McManus}, {McWilliams}, {Meacher},
  {Meadors}, {Meidam}, {Melatos}, {Mendell}, {Mendoza-Gandara}, {Mercer},
  {Merilh}, {Merzougui}, {Meshkov}, {Messenger}, {Messick}, {Meyers},
  {Mezzani}, {Miao}, {Michel}, {Middleton}, {Mikhailov}, {Milano}, {Miller},
  {Millhouse}, {Minenkov}, {Ming}, {Mirshekari}, {Mishra}, {Mitra},
  {Mitrofanov}, {Mitselmakher}, {Mittleman}, {Moggi}, {Mohan}, {Mohapatra},
  {Montani}, {Moore}, {Moore}, {Moraru}, {Moreno}, {Morriss}, {Mossavi},
  {Mours}, {Mow-Lowry}, {Mueller}, {Mueller}, {Muir}, {Mukherjee}, {Mukherjee},
  {Mukherjee}, {Mukund}, {Mullavey}, {Munch}, {Murphy}, {Murray}, {Mytidis},
  {Nardecchia}, {Naticchioni}, {Nayak}, {Necula}, {Nedkova}, {Nelemans},
  {Neri}, {Neunzert}, {Newton}, {Nguyen}, {Nielsen}, {Nissanke}, {Nitz},
  {Nocera}, {Nolting}, {Normandin}, {Nuttall}, {Oberling}, {Ochsner}, {O'Dell},
  {Oelker}, {Ogin}, {Oh}, {Oh}, {Ohme}, {Oliver}, {Oppermann}, {Oram},
  {O'Reilly}, {O'Shaughnessy}, {Ott}, {Ottaway}, {Ottens}, {Overmier}, {Owen},
  {Pai}, {Pai}, {Palamos}, {Palashov}, {Palomba}, {Pal-Singh}, {Pan}, {Pan},
  {Pankow}, {Pannarale}, {Pant}, {Paoletti}, {Paoli}, {Papa}, {Paris},
  {Parker}, {Pascucci}, {Pasqualetti}, {Passaquieti}, {Passuello},
  {Patricelli}, {Patrick}, {Pearlstone}, {Pedraza}, {Pedurand}, {Pekowsky},
  {Pele}, {Penn}, {Perreca}, {Pfeiffer}, {Phelps}, {Piccinni}, {Pichot},
  {Pickenpack}, {Piergiovanni}, {Pierro}, {Pillant}, {Pinard}, {Pinto},
  {Pitkin}, {Poeld}, {Poggiani}, {Popolizio}, {Post}, {Powell}, {Prasad},
  {Predoi}, {Premachandra}, {Prestegard}, {Price}, {Prijatelj}, {Principe},
  {Privitera}, {Prix}, {Prodi}, {Prokhorov}, {Puncken}, {Punturo}, {Puppo},
  {P{\"u}rrer}, {Qi}, {Qin}, {Quetschke}, {Quintero}, {Quitzow-James}, {Raab},
  {Rabeling}, {Radkins}, {Raffai}, {Raja}, {Rakhmanov}, {Ramet}, {Rapagnani},
  {Raymond}, {Razzano}, {Re}, {Read}, {Reed}, {Regimbau}, {Rei}, {Reid},
  {Reitze}, {Rew}, {Reyes}, {Ricci}, {Riles}, {Robertson}, {Robie}, {Robinet},
  {Rocchi}, {Rolland}, {Rollins}, {Roma}, {Romano}, {Romano}, {Romanov},
  {Romie}, {Rosi{\'n}ska}, {Rowan}, {R{\"u}diger}, {Ruggi}, {Ryan}, {Sachdev},
  {Sadecki}, {Sadeghian}, {Salconi}, {Saleem}, {Salemi}, {Samajdar}, {Sammut},
  {Sampson}, {Sanchez}, {Sandberg}, {Sandeen}, {Sanders}, {Sanders},
  {Sassolas}, {Sathyaprakash}, {Saulson}, {Sauter}, {Savage}, {Sawadsky},
  {Schale}, {Schilling}, {Schmidt}, {Schmidt}, {Schnabel}, {Schofield},
  {Sch{\"o}nbeck}, {Schreiber}, {Schuette}, {Schutz}, {Scott}, {Scott},
  {Sellers}, {Sengupta}, {Sentenac}, {Sequino}, {Sergeev}, {Serna},
  {Setyawati}, {Sevigny}, {Shaddock}, {Shaffer}, {Shah}, {Shahriar}, {Shaltev},
  {Shao}, {Shapiro}, {Shawhan}, {Sheperd}, {Shoemaker}, {Shoemaker}, {Siellez},
  {Siemens}, {Sigg}, {Silva}, {Simakov}, {Singer}, {Singer}, {Singh}, {Singh},
  {Singhal}, {Sintes}, {Slagmolen}, {Smith}, {Smith}, {Smith}, {Smith}, {Son},
  {Sorazu}, {Sorrentino}, {Souradeep}, {Srivastava}, {Staley}, {Steinke},
  {Steinlechner}, {Steinlechner}, {Steinmeyer}, {Stephens}, {Stevenson},
  {Stone}, {Strain}, {Straniero}, {Stratta}, {Strauss}, {Strigin}, {Sturani},
  {Stuver}, {Summerscales}, {Sun}, {Sutton}, {Swinkels}, {Szczepa{\'n}czyk},
  {Tacca}, {Talukder}, {Tanner}, {T{\'a}pai}, {Tarabrin}, {Taracchini},
  {Taylor}, {Theeg}, {Thirugnanasambandam}, {Thomas}, {Thomas}, {Thomas},
  {Thorne}, {Thorne}, {Thrane}, {Tiwari}, {Tiwari}, {Tokmakov}, {Tomlinson},
  {Tonelli}, {Torres}, {Torrie}, {T{\"o}yr{\"a}}, {Travasso}, {Traylor},
  {Trifir{\`o}}, {Tringali}, {Trozzo}, {Tse}, {Turconi}, {Tuyenbayev},
  {Ugolini}, {Unnikrishnan}, {Urban}, {Usman}, {Vahlbruch}, {Vajente},
  {Valdes}, {Vallisneri}, {van Bakel}, {van Beuzekom}, {van den Brand}, {Van
  Den Broeck}, {Vander-Hyde}, {van der Schaaf}, {van Heijningen}, {van Veggel},
  {Vardaro}, {Vass}, {Vas{\'u}th}, {Vaulin}, {Vecchio}, {Vedovato}, {Veitch},
  {Veitch}, {Venkateswara}, {Verkindt}, {Vetrano}, {Vicer{\'e}}, {Vinciguerra},
  {Vine}, {Vinet}, {Vitale}, {Vo}, {Vocca}, {Vorvick}, {Voss}, {Vousden},
  {Vyatchanin}, {Wade}, {Wade}, {Wade}, {Waldman}, {Walker}, {Wallace},
  {Walsh}, {Wang}, {Wang}, {Wang}, {Wang}, {Wang}, {Ward}, {Ward}, {Warner},
  {Was}, {Weaver}, {Wei}, {Weinert}, {Weinstein}, {Weiss}, {Welborn}, {Wen},
  {We{\ss}els}, {Westphal}, {Wette}, {Whelan}, {Whitcomb}, {White}, {Whiting},
  {Wiesner}, {Wilkinson}, {Willems}, {Williams}, {Williams}, {Williamson},
  {Willis}, {Willke}, {Wimmer}, {Winkelmann}, {Winkler}, {Wipf}, {Wiseman},
  {Wittel}, {Woan}, {Worden}, {Wright}, {Wu}, {Yablon}, {Yakushin}, {Yam},
  {Yamamoto}, {Yancey}, {Yap}, {Yu}, {Yvert}, {Zadro{\.Z}ny}, {Zangrando},
  {Zanolin}, {Zendri}, {Zevin}, {Zhang}, {Zhang}, {Zhang}, {Zhang}, {Zhao},
  {Zhou}, {Zhou}, {Zhu}, {Zucker}, {Zuraw}, {Zweizig}, {LIGO Scientific
  Collaboration}, \& {Virgo Collaboration}}]{2016PhRvL.116f1102A}
{Abbott}, B.~P., {Abbott}, R., {Abbott}, T.~D., {et~al.} 2016, \prl, 116,
  061102, \dodoi{10.1103/PhysRevLett.116.061102}

\bibitem[{{Aerts}(2019)}]{Aerts:2019}
{Aerts}, C. 2019, arXiv e-prints, arXiv:1912.12300.
\newblock \doarXiv{1912.12300}

\bibitem[{{Aerts} {et~al.}(2010){Aerts}, {Christensen-Dalsgaard}, \&
  {Kurtz}}]{Aerts:2010}
{Aerts}, C., {Christensen-Dalsgaard}, J., \& {Kurtz}, D.~W. 2010,
  {Asteroseismology}

\bibitem[{{Aerts} {et~al.}(2019){Aerts}, {Mathis}, \&
  {Rogers}}]{Aerts_araa:2019}
{Aerts}, C., {Mathis}, S., \& {Rogers}, T.~M. 2019, \araa, 57, 35,
  \dodoi{10.1146/annurev-astro-091918-104359}

\bibitem[{{Aerts} {et~al.}(2009){Aerts}, {Puls}, {Godart}, \&
  {Dupret}}]{2009A&A...508..409A}
{Aerts}, C., {Puls}, J., {Godart}, M., \& {Dupret}, M.~A. 2009, \aap, 508, 409,
  \dodoi{10.1051/0004-6361/200810471}

\bibitem[{{Augustson} \& {Mathis}(2019)}]{Augustson:2019}
{Augustson}, K.~C., \& {Mathis}, S. 2019, \apj, 874, 83,
  \dodoi{10.3847/1538-4357/ab0b3d}

\bibitem[{{Augustson} {et~al.}(2020){Augustson}, {Mathis}, \&
  {Astoul}}]{Augustson:2020}
{Augustson}, K.~C., {Mathis}, S., \& {Astoul}, A. 2020, \apj, 903, 90,
  \dodoi{10.3847/1538-4357/abba1c}

\bibitem[{{Bailey} {et~al.}(2015){Bailey}, {Nagayama}, {Loisel}, {Rochau},
  {Blancard}, {Colgan}, {Cosse}, {Faussurier}, {Fontes}, {Gilleron},
  {Golovkin}, {Hansen}, {Iglesias}, {Kilcrease}, {Macfarlane}, {Mancini},
  {Nahar}, {Orban}, {Pain}, {Pradhan}, {Sherrill}, \& {Wilson}}]{Bailey:2015}
{Bailey}, J.~E., {Nagayama}, T., {Loisel}, G.~P., {et~al.} 2015, \nat, 517, 56,
  \dodoi{10.1038/nature14048}

\bibitem[{{Blomme} {et~al.}(2011){Blomme}, {Mahy}, {Catala}, {Cuypers},
  {Gosset}, {Godart}, {Montalban}, {Ventura}, {Rauw}, {Morel}, {Degroote},
  {Aerts}, {Noels}, {Michel}, {Baudin}, {Baglin}, {Auvergne}, \&
  {Samadi}}]{Blomme:2011}
{Blomme}, R., {Mahy}, L., {Catala}, C., {et~al.} 2011, \aap, 533, A4,
  \dodoi{10.1051/0004-6361/201116949}

\bibitem[{{B{\"o}hm-Vitense}(1958)}]{1958ZA.....46..108B}
{B{\"o}hm-Vitense}, E. 1958, \zap, 46, 108

\bibitem[{{Bowman}(2020)}]{BowmanReview:2020}
{Bowman}, D.~M. 2020, arXiv e-prints, arXiv:2008.11162.
\newblock \doarXiv{2008.11162}

\bibitem[{{Bowman} {et~al.}(2020){Bowman}, {Burssens}, {Sim{\'o}n-D{\'\i}az},
  {Edelmann}, {Rogers}, {Horst}, {R{\"o}pke}, \& {Aerts}}]{Bowman:2020}
{Bowman}, D.~M., {Burssens}, S., {Sim{\'o}n-D{\'\i}az}, S., {et~al.} 2020,
  \aap, 640, A36, \dodoi{10.1051/0004-6361/202038224}

\bibitem[{{Bowman} {et~al.}(2019{\natexlab{a}}){Bowman}, {Burssens},
  {Pedersen}, {Johnston}, {Aerts}, {Buysschaert}, {Michielsen}, {Tkachenko},
  {Rogers}, {Edelmann}, {Ratnasingam}, {Sim{\'o}n-D{\'\i}az}, {Castro},
  {Moravveji}, {Pope}, {White}, \& {De Cat}}]{2019NatAs...3..760B}
{Bowman}, D.~M., {Burssens}, S., {Pedersen}, M.~G., {et~al.}
  2019{\natexlab{a}}, Nature Astronomy, 3, 760,
  \dodoi{10.1038/s41550-019-0768-1}

\bibitem[{{Bowman} {et~al.}(2019{\natexlab{b}}){Bowman}, {Aerts}, {Johnston},
  {Pedersen}, {Rogers}, {Edelmann}, {Sim{\'o}n-D{\'\i}az}, {Van Reeth},
  {Buysschaert}, {Tkachenko}, \& {Triana}}]{Bowman:2019}
{Bowman}, D.~M., {Aerts}, C., {Johnston}, C., {et~al.} 2019{\natexlab{b}},
  \aap, 621, A135, \dodoi{10.1051/0004-6361/201833662}

\bibitem[{{Buchler} \& {Yueh}(1976)}]{Buchler1976}
{Buchler}, J.~R., \& {Yueh}, W.~R. 1976, \apj, 210, 440, \dodoi{10.1086/154847}

\bibitem[{{Burssens} {et~al.}(2020){Burssens}, {Sim{\'o}n-D{\'\i}az}, {Bowman},
  {Holgado}, {Michielsen}, {de Burgos}, {Castro}, {Barb{\'a}}, \&
  {Aerts}}]{Burssens:2020}
{Burssens}, S., {Sim{\'o}n-D{\'\i}az}, S., {Bowman}, D.~M., {et~al.} 2020,
  \aap, 639, A81, \dodoi{10.1051/0004-6361/202037700}

\bibitem[{{Cantiello} \& {Braithwaite}(2011)}]{Cantiello:2011}
{Cantiello}, M., \& {Braithwaite}, J. 2011, \aap, 534, A140,
  \dodoi{10.1051/0004-6361/201117512}

\bibitem[{{Cantiello} \& {Braithwaite}(2019)}]{Cantiello:2019}
---. 2019, \apj, 883, 106, \dodoi{10.3847/1538-4357/ab3924}

\bibitem[{{Cantiello} {et~al.}(2009){Cantiello}, {Langer}, {Brott}, {de Koter},
  {Shore}, {Vink}, {Voegler}, {Lennon}, \& {Yoon}}]{Cantiello:2009}
{Cantiello}, M., {Langer}, N., {Brott}, I., {et~al.} 2009, \aap, 499, 279,
  \dodoi{10.1051/0004-6361/200911643}

\bibitem[{{Cassisi} {et~al.}(2007){Cassisi}, {Potekhin}, {Pietrinferni},
  {Catelan}, \& {Salaris}}]{Cassisi2007}
{Cassisi}, S., {Potekhin}, A.~Y., {Pietrinferni}, A., {Catelan}, M., \&
  {Salaris}, M. 2007, \apj, 661, 1094, \dodoi{10.1086/516819}

\bibitem[{{Castro} {et~al.}(2015){Castro}, {Fossati}, {Hubrig},
  {Sim{\'o}n-D{\'\i}az}, {Sch{\"o}ller}, {Ilyin}, {Carrol}, {Langer}, {Morel},
  {Schneider}, {Przybilla}, {Herrero}, {de Koter}, {Oskinova}, {Reisenegger},
  {Sana}, \& {BOB Collaboration}}]{2015A&A...581A..81C}
{Castro}, N., {Fossati}, L., {Hubrig}, S., {et~al.} 2015, \aap, 581, A81,
  \dodoi{10.1051/0004-6361/201425354}

\bibitem[{{Chugunov} {et~al.}(2007){Chugunov}, {Dewitt}, \&
  {Yakovlev}}]{Chugunov2007}
{Chugunov}, A.~I., {Dewitt}, H.~E., \& {Yakovlev}, D.~G. 2007, \prd, 76,
  025028, \dodoi{10.1103/PhysRevD.76.025028}

\bibitem[{{Cranmer} \& {Owocki}(1996)}]{Cranmer:1996}
{Cranmer}, S.~R., \& {Owocki}, S.~P. 1996, \apj, 462, 469,
  \dodoi{10.1086/177166}

\bibitem[{{Cyburt} {et~al.}(2010){Cyburt}, {Amthor}, {Ferguson}, {Meisel},
  {Smith}, {Warren}, {Heger}, {Hoffman}, {Rauscher}, {Sakharuk}, {Schatz},
  {Thielemann}, \& {Wiescher}}]{Cyburt2010}
{Cyburt}, R.~H., {Amthor}, A.~M., {Ferguson}, R., {et~al.} 2010, \apjs, 189,
  240, \dodoi{10.1088/0067-0049/189/1/240}

\bibitem[{{de Mink} {et~al.}(2014){de Mink}, {Sana}, {Langer}, {Izzard}, \&
  {Schneider}}]{deMink:2014}
{de Mink}, S.~E., {Sana}, H., {Langer}, N., {Izzard}, R.~G., \& {Schneider},
  F.~R.~N. 2014, \apj, 782, 7, \dodoi{10.1088/0004-637X/782/1/7}

\bibitem[{{Dintrans} \& {Rieutord}(2000)}]{Dintrans:2000}
{Dintrans}, B., \& {Rieutord}, M. 2000, \aap, 354, 86

\bibitem[{{Dorn-Wallenstein} {et~al.}(2020){Dorn-Wallenstein}, {Levesque},
  {Neugent}, {Davenport}, {Morris}, \& {Gootkin}}]{Dorn:2020}
{Dorn-Wallenstein}, T.~Z., {Levesque}, E.~M., {Neugent}, K.~F., {et~al.} 2020,
  \apj, 902, 24, \dodoi{10.3847/1538-4357/abb318}

\bibitem[{{Dufton} {et~al.}(2013){Dufton}, {Langer}, {Dunstall}, {Evans},
  {Brott}, {de Mink}, {Howarth}, {Kennedy}, {McEvoy}, {Potter},
  {Ram{\'\i}rez-Agudelo}, {Sana}, {Sim{\'o}n-D{\'\i}az}, {Taylor}, \&
  {Vink}}]{Dufton:2013}
{Dufton}, P.~L., {Langer}, N., {Dunstall}, P.~R., {et~al.} 2013, \aap, 550,
  A109, \dodoi{10.1051/0004-6361/201220273}

\bibitem[{{Dziembowski}(1977)}]{Dziembowski:1977}
{Dziembowski}, W. 1977, \actaa, 27, 203

\bibitem[{{Edelmann} {et~al.}(2019){Edelmann}, {Ratnasingam}, {Pedersen},
  {Bowman}, {Prat}, \& {Rogers}}]{Edelmann:2019}
{Edelmann}, P.~V.~F., {Ratnasingam}, R.~P., {Pedersen}, M.~G., {et~al.} 2019,
  \apj, 876, 4, \dodoi{10.3847/1538-4357/ab12df}

\bibitem[{{Ferguson} {et~al.}(2005){Ferguson}, {Alexander}, {Allard}, {Barman},
  {Bodnarik}, {Hauschildt}, {Heffner-Wong}, \& {Tamanai}}]{Ferguson2005}
{Ferguson}, J.~W., {Alexander}, D.~R., {Allard}, F., {et~al.} 2005, \apj, 623,
  585, \dodoi{10.1086/428642}

\bibitem[{{Fuller} {et~al.}(1985){Fuller}, {Fowler}, \& {Newman}}]{Fuller1985}
{Fuller}, G.~M., {Fowler}, W.~A., \& {Newman}, M.~J. 1985, \apj, 293, 1,
  \dodoi{10.1086/163208}

\bibitem[{{Fuller} {et~al.}(2015){Fuller}, {Cantiello}, {Stello}, {Garcia}, \&
  {Bildsten}}]{2015Sci...350..423F}
{Fuller}, J., {Cantiello}, M., {Stello}, D., {Garcia}, R.~A., \& {Bildsten}, L.
  2015, Science, 350, 423, \dodoi{10.1126/science.aac6933}

\bibitem[{{Fullerton} {et~al.}(1996){Fullerton}, {Gies}, \&
  {Bolton}}]{Fullerton:1996}
{Fullerton}, A.~W., {Gies}, D.~R., \& {Bolton}, C.~T. 1996, \apjs, 103, 475,
  \dodoi{10.1086/192285}

\bibitem[{{Fullerton} {et~al.}(1997){Fullerton}, {Massa}, {Prinja}, {Owocki},
  \& {Cranmer}}]{Fullerton:1997}
{Fullerton}, A.~W., {Massa}, D.~L., {Prinja}, R.~K., {Owocki}, S.~P., \&
  {Cranmer}, S.~R. 1997, \aap, 327, 699

\bibitem[{{Godart} {et~al.}(2017){Godart}, {Sim{\'o}n-D{\'\i}az}, {Herrero},
  {Dupret}, {Gr{\"o}tsch-Noels}, {Salmon}, \& {Ventura}}]{Godart:2017}
{Godart}, M., {Sim{\'o}n-D{\'\i}az}, S., {Herrero}, A., {et~al.} 2017, \aap,
  597, A23, \dodoi{10.1051/0004-6361/201628856}

\bibitem[{{Goldreich} \& {Kumar}(1990)}]{1990ApJ...363..694G}
{Goldreich}, P., \& {Kumar}, P. 1990, \apj, 363, 694, \dodoi{10.1086/169376}

\bibitem[{{Grassitelli} {et~al.}(2015{\natexlab{a}}){Grassitelli}, {Fossati},
  {Langer}, {Miglio}, {Istrate}, \& {Sanyal}}]{Grassitelli:2015b}
{Grassitelli}, L., {Fossati}, L., {Langer}, N., {et~al.} 2015{\natexlab{a}},
  \aap, 584, L2, \dodoi{10.1051/0004-6361/201527289}

\bibitem[{{Grassitelli} {et~al.}(2016){Grassitelli}, {Fossati}, {Langer},
  {Sim{\'o}n-D{\'\i}az}, {Castro}, \& {Sanyal}}]{Grassitelli:2016}
---. 2016, \aap, 593, A14, \dodoi{10.1051/0004-6361/201628912}

\bibitem[{{Grassitelli} {et~al.}(2015{\natexlab{b}}){Grassitelli}, {Fossati},
  {Sim{\'o}n-Di{\'a}z}, {Langer}, {Castro}, \& {Sanyal}}]{Grassitelli:2015}
{Grassitelli}, L., {Fossati}, L., {Sim{\'o}n-Di{\'a}z}, S., {et~al.}
  2015{\natexlab{b}}, \apjl, 808, L31, \dodoi{10.1088/2041-8205/808/1/L31}

\bibitem[{{Grevesse} \& {Sauval}(1998)}]{1998SSRv...85..161G}
{Grevesse}, N., \& {Sauval}, A.~J. 1998, \ssr, 85, 161,
  \dodoi{10.1023/A:1005161325181}

\bibitem[{{Hopkins} {et~al.}(2014){Hopkins}, {Kere{\v{s}}}, {O{\~n}orbe},
  {Faucher-Gigu{\`e}re}, {Quataert}, {Murray}, \& {Bullock}}]{Hopkins:2014}
{Hopkins}, P.~F., {Kere{\v{s}}}, D., {O{\~n}orbe}, J., {et~al.} 2014, \mnras,
  445, 581, \dodoi{10.1093/mnras/stu1738}

\bibitem[{{Horst} {et~al.}(2020){Horst}, {Edelmann}, {Andr{\'a}ssy},
  {R{\"o}pke}, {Bowman}, {Aerts}, \& {Ratnasingam}}]{Horst:2020}
{Horst}, L., {Edelmann}, P.~V.~F., {Andr{\'a}ssy}, R., {et~al.} 2020, \aap,
  641, A18, \dodoi{10.1051/0004-6361/202037531}

\bibitem[{{Howarth} \& {Prinja}(1989)}]{Howarth:1989}
{Howarth}, I.~D., \& {Prinja}, R.~K. 1989, \apjs, 69, 527,
  \dodoi{10.1086/191321}

\bibitem[{{Iglesias} \& {Rogers}(1993)}]{Iglesias1993}
{Iglesias}, C.~A., \& {Rogers}, F.~J. 1993, \apj, 412, 752,
  \dodoi{10.1086/172958}

\bibitem[{{Iglesias} \& {Rogers}(1996)}]{Iglesias1996}
---. 1996, \apj, 464, 943, \dodoi{10.1086/177381}

\bibitem[{{Irwin}(2004)}]{Irwin2004}
{Irwin}, A.~W. 2004, The FreeEOS Code for Calculating the Equation of State for
  Stellar Interiors.
\newblock \url{http://freeeos.sourceforge.net/}

\bibitem[{{Itoh} {et~al.}(1996){Itoh}, {Hayashi}, {Nishikawa}, \&
  {Kohyama}}]{Itoh1996}
{Itoh}, N., {Hayashi}, H., {Nishikawa}, A., \& {Kohyama}, Y. 1996, \apjs, 102,
  411, \dodoi{10.1086/192264}

\bibitem[{{Jermyn} \& {Cantiello}(2020)}]{Jermyn:2020}
{Jermyn}, A.~S., \& {Cantiello}, M. 2020, \apj, 900, 113,
  \dodoi{10.3847/1538-4357/ab9e70}

\bibitem[{{Jiang} {et~al.}(2015){Jiang}, {Cantiello}, {Bildsten}, {Quataert},
  \& {Blaes}}]{2015ApJ...813...74J}
{Jiang}, Y.-F., {Cantiello}, M., {Bildsten}, L., {Quataert}, E., \& {Blaes}, O.
  2015, \apj, 813, 74, \dodoi{10.1088/0004-637X/813/1/74}

\bibitem[{{Jiang} {et~al.}(2017){Jiang}, {Cantiello}, {Bildsten}, {Quataert},
  \& {Blaes}}]{Jiang:2017}
---. 2017, \apj, 843, 68, \dodoi{10.3847/1538-4357/aa77b0}

\bibitem[{{Jiang} {et~al.}(2018){Jiang}, {Cantiello}, {Bildsten}, {Quataert},
  {Blaes}, \& {Stone}}]{2018Natur.561..498J}
{Jiang}, Y.-F., {Cantiello}, M., {Bildsten}, L., {et~al.} 2018, \nat, 561, 498,
  \dodoi{10.1038/s41586-018-0525-0}

\bibitem[{{Kallinger} {et~al.}(2014){Kallinger}, {De Ridder}, {Hekker},
  {Mathur}, {Mosser}, {Gruberbauer}, {Garc{\'\i}a}, {Karoff}, \&
  {Ballot}}]{Kallinger:2014}
{Kallinger}, T., {De Ridder}, J., {Hekker}, S., {et~al.} 2014, \aap, 570, A41,
  \dodoi{10.1051/0004-6361/201424313}

\bibitem[{{Kaper} {et~al.}(1997){Kaper}, {Henrichs}, {Fullerton}, {Ando},
  {Bjorkman}, {Gies}, {Hirata}, {Kambe}, {McDavid}, \& {Nichols}}]{Kaper:1997}
{Kaper}, L., {Henrichs}, H.~F., {Fullerton}, A.~W., {et~al.} 1997, \aap, 327,
  281

\bibitem[{{Kjeldsen} \& {Bedding}(1995)}]{Kjeldsen:1995}
{Kjeldsen}, H., \& {Bedding}, T.~R. 1995, \aap, 293, 87.
\newblock \doarXiv{astro-ph/9403015}

\bibitem[{{Krticka} \& {Feldmeier}(2021)}]{Krticka:2021}
{Krticka}, J., \& {Feldmeier}, A. 2021, arXiv e-prints, arXiv:2103.08755.
\newblock \doarXiv{2103.08755}

\bibitem[{{Landstreet} {et~al.}(1989){Landstreet}, {Barker}, {Bohlender}, \&
  {Jewison}}]{1989ApJ...344..876L}
{Landstreet}, J.~D., {Barker}, P.~K., {Bohlender}, D.~A., \& {Jewison}, M.~S.
  1989, \apj, 344, 876, \dodoi{10.1086/167855}

\bibitem[{{Langanke} \& {Mart{\'{\i}}nez-Pinedo}(2000)}]{Langanke2000}
{Langanke}, K., \& {Mart{\'{\i}}nez-Pinedo}, G. 2000, Nuclear Physics A, 673,
  481, \dodoi{10.1016/S0375-9474(00)00131-7}

\bibitem[{{Langer}(2012)}]{Langer:2012}
{Langer}, N. 2012, \araa, 50, 107, \dodoi{10.1146/annurev-astro-081811-125534}

\bibitem[{{Langer} \& {Kudritzki}(2014)}]{2014A&A...564A..52L}
{Langer}, N., \& {Kudritzki}, R.~P. 2014, \aap, 564, A52,
  \dodoi{10.1051/0004-6361/201423374}

\bibitem[{{Lecoanet} {et~al.}(2021){Lecoanet}, {Cantiello}, {Anders},
  {Quataert}, {Couston}, {Bouffard}, {Favier}, \& {Le Bars}}]{Lecoanet:2021}
{Lecoanet}, D., {Cantiello}, M., {Anders}, E.~H., {et~al.} 2021, arXiv
  e-prints, arXiv:2105.04558.
\newblock \doarXiv{2105.04558}

\bibitem[{{Lecoanet} \& {Quataert}(2013)}]{2013MNRAS.430.2363L}
{Lecoanet}, D., \& {Quataert}, E. 2013, \mnras, 430, 2363,
  \dodoi{10.1093/mnras/stt055}

\bibitem[{{Lecoanet} {et~al.}(2019){Lecoanet}, {Cantiello}, {Quataert},
  {Couston}, {Burns}, {Pope}, {Jermyn}, {Favier}, \& {Le Bars}}]{Lecoanet:2020}
{Lecoanet}, D., {Cantiello}, M., {Quataert}, E., {et~al.} 2019, \apjl, 886,
  L15, \dodoi{10.3847/2041-8213/ab5446}

\bibitem[{{MacDonald} \& {Petit}(2019)}]{2019MNRAS.487.3904M}
{MacDonald}, J., \& {Petit}, V. 2019, \mnras, 487, 3904,
  \dodoi{10.1093/mnras/stz1545}

\bibitem[{{Maeder} {et~al.}(2008){Maeder}, {Georgy}, \& {Meynet}}]{Maeder:2008}
{Maeder}, A., {Georgy}, C., \& {Meynet}, G. 2008, \aap, 479, L37,
  \dodoi{10.1051/0004-6361:20079007}

\bibitem[{{Maeder} \& {Meynet}(2000)}]{Maeder:2000}
{Maeder}, A., \& {Meynet}, G. 2000, \araa, 38, 143,
  \dodoi{10.1146/annurev.astro.38.1.143}

\bibitem[{{Mathis} {et~al.}(2014){Mathis}, {Neiner}, \& {Tran
  Minh}}]{Mathis:2014}
{Mathis}, S., {Neiner}, C., \& {Tran Minh}, N. 2014, \aap, 565, A47,
  \dodoi{10.1051/0004-6361/201321830}

\bibitem[{{Naze} {et~al.}(2021){Naze}, {Rauw}, \& {Gosset}}]{Naze:2021}
{Naze}, Y., {Rauw}, G., \& {Gosset}, E. 2021, arXiv e-prints, arXiv:2101.05521.
\newblock \doarXiv{2101.05521}

\bibitem[{{Neiner} {et~al.}(2020){Neiner}, {Lee}, {Mathis}, {Saio}, {Lovekin},
  \& {Augustson}}]{Neiner:2020}
{Neiner}, C., {Lee}, U., {Mathis}, S., {et~al.} 2020, \aap, 644, A9,
  \dodoi{10.1051/0004-6361/201935858}

\bibitem[{{Oda} {et~al.}(1994){Oda}, {Hino}, {Muto}, {Takahara}, \&
  {Sato}}]{Oda1994}
{Oda}, T., {Hino}, M., {Muto}, K., {Takahara}, M., \& {Sato}, K. 1994, Atomic
  Data and Nuclear Data Tables, 56, 231, \dodoi{10.1006/adnd.1994.1007}

\bibitem[{{Owocki} {et~al.}(1988){Owocki}, {Castor}, \&
  {Rybicki}}]{Owocki:1988}
{Owocki}, S.~P., {Castor}, J.~I., \& {Rybicki}, G.~B. 1988, \apj, 335, 914,
  \dodoi{10.1086/166977}

\bibitem[{{Paxton} {et~al.}(2011){Paxton}, {Bildsten}, {Dotter}, {Herwig},
  {Lesaffre}, \& {Timmes}}]{Paxton2011}
{Paxton}, B., {Bildsten}, L., {Dotter}, A., {et~al.} 2011, \apjs, 192, 3,
  \dodoi{10.1088/0067-0049/192/1/3}

\bibitem[{{Paxton} {et~al.}(2013){Paxton}, {Cantiello}, {Arras}, {Bildsten},
  {Brown}, {Dotter}, {Mankovich}, {Montgomery}, {Stello}, {Timmes}, \&
  {Townsend}}]{Paxton2013}
{Paxton}, B., {Cantiello}, M., {Arras}, P., {et~al.} 2013, \apjs, 208, 4,
  \dodoi{10.1088/0067-0049/208/1/4}

\bibitem[{{Paxton} {et~al.}(2015){Paxton}, {Marchant}, {Schwab}, {Bauer},
  {Bildsten}, {Cantiello}, {Dessart}, {Farmer}, {Hu}, {Langer}, {Townsend},
  {Townsley}, \& {Timmes}}]{Paxton2015}
{Paxton}, B., {Marchant}, P., {Schwab}, J., {et~al.} 2015, \apjs, 220, 15,
  \dodoi{10.1088/0067-0049/220/1/15}

\bibitem[{{Paxton} {et~al.}(2018){Paxton}, {Schwab}, {Bauer}, {Bildsten},
  {Blinnikov}, {Duffell}, {Farmer}, {Goldberg}, {Marchant}, {Sorokina},
  {Thoul}, {Townsend}, \& {Timmes}}]{Paxton2018}
{Paxton}, B., {Schwab}, J., {Bauer}, E.~B., {et~al.} 2018, \apjs, 234, 34,
  \dodoi{10.3847/1538-4365/aaa5a8}

\bibitem[{{Paxton} {et~al.}(2019){Paxton}, {Smolec}, {Schwab}, {Gautschy},
  {Bildsten}, {Cantiello}, {Dotter}, {Farmer}, {Goldberg}, {Jermyn}, {Kanbur},
  {Marchant}, {Thoul}, {Townsend}, {Wolf}, {Zhang}, \& {Timmes}}]{Paxton2019}
{Paxton}, B., {Smolec}, R., {Schwab}, J., {et~al.} 2019, \apjs, 243, 10,
  \dodoi{10.3847/1538-4365/ab2241}

\bibitem[{{Pedersen} {et~al.}(2019){Pedersen}, {Chowdhury}, {Johnston},
  {Bowman}, {Aerts}, {Handler}, {De Cat}, {Neiner}, {David-Uraz}, {Buzasi},
  {Tkachenko}, {Sim{\'o}n-D{\'\i}az}, {Moravveji}, {Sikora}, {Mirouh},
  {Lovekin}, {Cantiello}, {Daszy{\'n}ska-Daszkiewicz}, {Pigulski},
  {Vanderspek}, \& {Ricker}}]{Pedersen:2019}
{Pedersen}, M.~G., {Chowdhury}, S., {Johnston}, C., {et~al.} 2019, \apjl, 872,
  L9, \dodoi{10.3847/2041-8213/ab01e1}

\bibitem[{{Penny} \& {Gies}(2009)}]{Penny:2009}
{Penny}, L.~R., \& {Gies}, D.~R. 2009, \apj, 700, 844,
  \dodoi{10.1088/0004-637X/700/1/844}

\bibitem[{{Potekhin} \& {Chabrier}(2010)}]{Potekhin2010}
{Potekhin}, A.~Y., \& {Chabrier}, G. 2010, Contributions to Plasma Physics, 50,
  82, \dodoi{10.1002/ctpp.201010017}

\bibitem[{{Prat} {et~al.}(2016){Prat}, {Ligni{\`e}res}, \&
  {Ballot}}]{Prat:2016}
{Prat}, V., {Ligni{\`e}res}, F., \& {Ballot}, J. 2016, \aap, 587, A110,
  \dodoi{10.1051/0004-6361/201527737}

\bibitem[{{Press}(1981)}]{Press:1981}
{Press}, W.~H. 1981, \apj, 245, 286, \dodoi{10.1086/158809}

\bibitem[{{Ramiaramanantsoa} {et~al.}(2014){Ramiaramanantsoa}, {Moffat},
  {Chen{\'e}}, {Richardson}, {Henrichs}, {Desforges}, {Antoci}, {Rowe},
  {Matthews}, {Kuschnig}, {Weiss}, {Sasselov}, {Rucinski}, \&
  {Guenther}}]{Ramiaramanantsoa:2014}
{Ramiaramanantsoa}, T., {Moffat}, A.~F.~J., {Chen{\'e}}, A.-N., {et~al.} 2014,
  \mnras, 441, 910, \dodoi{10.1093/mnras/stu619}

\bibitem[{{Ram{\'\i}rez-Agudelo} {et~al.}(2013){Ram{\'\i}rez-Agudelo},
  {Sim{\'o}n-D{\'\i}az}, {Sana}, {de Koter}, {Sab{\'\i}n-Sanjul{\'\i}an}, {de
  Mink}, {Dufton}, {Gr{\"a}fener}, {Evans}, {Herrero}, {Langer}, {Lennon},
  {Ma{\'\i}z Apell{\'a}niz}, {Markova}, {Najarro}, {Puls}, {Taylor}, \&
  {Vink}}]{Ramirez:2013}
{Ram{\'\i}rez-Agudelo}, O.~H., {Sim{\'o}n-D{\'\i}az}, S., {Sana}, H., {et~al.}
  2013, \aap, 560, A29, \dodoi{10.1051/0004-6361/201321986}

\bibitem[{{Ratnasingam} {et~al.}(2019){Ratnasingam}, {Edelmann}, \&
  {Rogers}}]{Ratnasingam:2019}
{Ratnasingam}, R.~P., {Edelmann}, P.~V.~F., \& {Rogers}, T.~M. 2019, \mnras,
  482, 5500, \dodoi{10.1093/mnras/sty3086}

\bibitem[{{Ratnasingam} {et~al.}(2020){Ratnasingam}, {Edelmann}, \&
  {Rogers}}]{Ratnasingam:2020}
---. 2020, \mnras, \dodoi{10.1093/mnras/staa2296}

\bibitem[{{Rauw} \& {Naz{\'e}}(2021)}]{Rauw:2021}
{Rauw}, G., \& {Naz{\'e}}, Y. 2021, \mnras, 500, 2096,
  \dodoi{10.1093/mnras/staa3310}

\bibitem[{{Richer} {et~al.}(2000){Richer}, {Michaud}, \&
  {Turcotte}}]{Richer:2000}
{Richer}, J., {Michaud}, G., \& {Turcotte}, S. 2000, \apj, 529, 338,
  \dodoi{10.1086/308274}

\bibitem[{{Rogers} \& {Nayfonov}(2002)}]{Rogers2002}
{Rogers}, F.~J., \& {Nayfonov}, A. 2002, \apj, 576, 1064,
  \dodoi{10.1086/341894}

\bibitem[{{Rogers} {et~al.}(2013){Rogers}, {Lin}, {McElwaine}, \&
  {Lau}}]{2013ApJ...772...21R}
{Rogers}, T.~M., {Lin}, D.~N.~C., {McElwaine}, J.~N., \& {Lau}, H.~H.~B. 2013,
  \apj, 772, 21, \dodoi{10.1088/0004-637X/772/1/21}

\bibitem[{{Sana} {et~al.}(2012){Sana}, {de Mink}, {de Koter}, {Langer},
  {Evans}, {Gieles}, {Gosset}, {Izzard}, {Le Bouquin}, \&
  {Schneider}}]{Sana:2012}
{Sana}, H., {de Mink}, S.~E., {de Koter}, A., {et~al.} 2012, Science, 337, 444,
  \dodoi{10.1126/science.1223344}

\bibitem[{{Saumon} {et~al.}(1995){Saumon}, {Chabrier}, \& {van
  Horn}}]{Saumon1995}
{Saumon}, D., {Chabrier}, G., \& {van Horn}, H.~M. 1995, \apjs, 99, 713,
  \dodoi{10.1086/192204}

\bibitem[{{Schultz} {et~al.}(2020){Schultz}, {Bildsten}, \&
  {Jiang}}]{Schultz:2020}
{Schultz}, W.~C., {Bildsten}, L., \& {Jiang}, Y.-F. 2020, \apj, 902, 67,
  \dodoi{10.3847/1538-4357/abb405}

\bibitem[{{Shiode} {et~al.}(2013){Shiode}, {Quataert}, {Cantiello}, \&
  {Bildsten}}]{Shiode:2013}
{Shiode}, J.~H., {Quataert}, E., {Cantiello}, M., \& {Bildsten}, L. 2013,
  \mnras, 430, 1736, \dodoi{10.1093/mnras/sts719}

\bibitem[{{Sim{\'o}n-D{\'\i}az}
  {et~al.}(2017{\natexlab{a}}){Sim{\'o}n-D{\'\i}az}, {Godart}, {Castro},
  {Herrero}, {Aerts}, {Puls}, {Telting}, \& {Grassitelli}}]{Simon-Diaz:2017}
{Sim{\'o}n-D{\'\i}az}, S., {Godart}, M., {Castro}, N., {et~al.}
  2017{\natexlab{a}}, \aap, 597, A22, \dodoi{10.1051/0004-6361/201628541}

\bibitem[{{Sim{\'o}n-D{\'\i}az}
  {et~al.}(2017{\natexlab{b}}){Sim{\'o}n-D{\'\i}az}, {Godart}, {Castro},
  {Herrero}, {Aerts}, {Puls}, {Telting}, \&
  {Grassitelli}}]{2017A&A...597A..22S}
---. 2017{\natexlab{b}}, \aap, 597, A22, \dodoi{10.1051/0004-6361/201628541}

\bibitem[{{Sim{\'o}n-D{\'\i}az} \& {Herrero}(2014)}]{Simon-Diaz:2014}
{Sim{\'o}n-D{\'\i}az}, S., \& {Herrero}, A. 2014, \aap, 562, A135,
  \dodoi{10.1051/0004-6361/201322758}

\bibitem[{{Smith}(2014)}]{Smith:2014}
{Smith}, N. 2014, \araa, 52, 487, \dodoi{10.1146/annurev-astro-081913-040025}

\bibitem[{{Stevenson}(1979)}]{Stevenson:1979}
{Stevenson}, D.~J. 1979, Geophysical and Astrophysical Fluid Dynamics, 12, 139,
  \dodoi{10.1080/03091927908242681}

\bibitem[{{Sudnik} \& {Henrichs}(2016)}]{Sudnik:2016}
{Sudnik}, N.~P., \& {Henrichs}, H.~F. 2016, \aap, 594, A56,
  \dodoi{10.1051/0004-6361/201628529}

\bibitem[{{Sundqvist} \& {Owocki}(2013)}]{Sundqvist:2013}
{Sundqvist}, J.~O., \& {Owocki}, S.~P. 2013, \mnras, 428, 1837,
  \dodoi{10.1093/mnras/sts165}

\bibitem[{Sundqvist {et~al.}(2013)Sundqvist, Petit, Owocki, Wade, Puls, \&
  Collaboration}]{10.1093/mnras/stt921}
Sundqvist, J.~O., Petit, V., Owocki, S.~P., {et~al.} 2013, Monthly Notices of
  the Royal Astronomical Society, 433, 2497, \dodoi{10.1093/mnras/stt921}

\bibitem[{Sutherland(2010)}]{sutherland_2010}
Sutherland, B.~R. 2010, Internal Gravity Waves (Cambridge University Press),
  \dodoi{10.1017/CBO9780511780318}

\bibitem[{{Szewczuk} {et~al.}(2021){Szewczuk}, {Walczak}, \&
  {Daszy{\'n}ska-Daszkiewicz}}]{Szewczuk:2021}
{Szewczuk}, W., {Walczak}, P., \& {Daszy{\'n}ska-Daszkiewicz}, J. 2021, \mnras,
  503, 5894, \dodoi{10.1093/mnras/stab683}

\bibitem[{{Timmes} \& {Swesty}(2000)}]{Timmes2000}
{Timmes}, F.~X., \& {Swesty}, F.~D. 2000, \apjs, 126, 501,
  \dodoi{10.1086/313304}

\bibitem[{{Yusof} {et~al.}(2013){Yusof}, {Hirschi}, {Meynet}, {Crowther},
  {Ekstr{\"o}m}, {Frischknecht}, {Georgy}, {Abu Kassim}, \&
  {Schnurr}}]{2013MNRAS.433.1114Y}
{Yusof}, N., {Hirschi}, R., {Meynet}, G., {et~al.} 2013, \mnras, 433, 1114,
  \dodoi{10.1093/mnras/stt794}

\end{thebibliography}

\end{document}